\documentclass[preprint,superscriptaddress,nofootinbib]{revtex4-1}
\usepackage[dvips]{graphicx}
\usepackage{amsmath,amssymb,color,colortbl}

\allowdisplaybreaks[4]

\newcommand{\red}[1]{{\color[rgb]{1,0,0} #1}}

\newcommand{\MNS}{{\text{MNS}}}
\newcommand{\DM}{{\text{DM}}}
\newcommand{\Pl}{{\text{Pl}}}
\newcommand{\SI}{{\text{SI}}}
\newcommand{\EM}{{\text{EM}}}
\newcommand{\eff}{{\text{eff}}}
\newcommand{\tree}{{\text{tree}}}
\newcommand{\rel}{{\text{rel}}}

\newcommand{\eV}{{\text{eV}}}

\newcommand{\GeV}{{\text{GeV}}}
\newcommand{\TeV}{{\text{TeV}}}

\newcommand{\fb}{{\text{fb}}}

\newcommand{\BR}{\text{BR}}

\newcommand{\U}{{\text{U}}}
\newcommand{\SU}{{\text{SU}}}

\newcommand{\ellp}{\ell^\prime}
\newcommand{\dprime}{{\prime\prime}}
\newcommand{\wt}[1]{\widetilde{#1}}

\newcommand{\LN}{L}
\newcommand{\BN}{B}

\begin{document}

\preprint{OU-HET 932}
\preprint{UT-HET 121}

\title{
Neutrino Mass,
Dark Matter and Baryon Asymmetry
without Lepton Number Violation
}

\author{Shinya Kanemura}
\email{kanemu@het.phys.sci.osaka-u.ac.jp}
\affiliation{
Department of Physics,
Osaka University,
Toyonaka,
Osaka 560-0043, Japan
}
\author{Kodai Sakurai}
\email{sakurai@jodo.sci.u-toyama.ac.jp}
\affiliation{
Department of Physics,
University of Toyama,
3190 Gofuku,
Toyama 930-8555, Japan
}
\author{Hiroaki Sugiyama}
\email{shiro324@gmail.com}
\affiliation{
Liberal Arts and Sciences,
Toyama Prefectural University,
Toyama 939-0398, Japan
}


\begin{abstract}

 We propose a model
to explain tiny masses of neutrinos
with the lepton number conservation,
where neither too heavy particles beyond the TeV-scale
nor tiny coupling constants are required.
 Assignments of conserving lepton numbers to new fields
result in an unbroken $Z_2$ symmetry
that stabilizes the dark matter candidate%
~(the lightest $Z_2$-odd particle).
 In this model,
$Z_2$-odd particles play an important role
to generate the mass of neutrinos. 
 The scalar dark matter in our model
can satisfy constraints on the dark matter abundance
and those from direct searches.
 It is also shown that the strong first-order phase transition,
which is required for the electroweak baryogenesis,
can be realized in our model.
 In addition,
the scalar potential can in principle contain CP-violating phases,
which can also be utilized for the baryogenesis.
 Therefore,
three problems in the standard model,
namely absence of neutrino masses,
the dark matter candidate,
and the mechanism to generate baryon asymmetry of the Universe,
may be simultaneously resolved at the TeV-scale.
 Phenomenology of this model is also discussed briefly.

\end{abstract}

\maketitle

\section{Introduction}

 The standard model~(SM) of particle physics
was established by the discovery of a Higgs boson
at the CERN large hadron collider~(LHC)~\cite{ref:2012Jul}.
 This does not mean that
theoretical particle physics has been completed
because several problems remain in the high energy physics.
 For example,
neutrinos are massless in the SM
while the discovery of neutrino oscillations~\cite{ref:numass}
is the evidence of tiny neutrino masses.
 The absence of the candidate for the dark matter in the SM,
which accounts for 27\% of the Universe~\cite{Ade:2015xua},
must be resolved.
 The baryon asymmetry of the Universe
cannot be explained in the SM,
so that the SM must be extended
to include a mechanism to generate the baryon asymmetry of the Universe.

 Reasons why neutrinos are massless in the SM
are the absence of right-handed neutrinos $\nu_R$
and the lepton number conservation.
 If we introduce $\nu_R$ to the SM,
neutrino masses can be generated
via the naive Yukawa interaction
with the Higgs doublet field in the SM
similarly to masses of quarks and charged leptons.
 However,
the Yukawa coupling constant
seems to be unnaturally small~($\lesssim 10^{-12}$).
 Instead of such fine-tuned coupling constants,
in the (type-I) seesaw mechanism~\cite{ref:seesaw}
very large Majorana masses of $\nu_R$
are introduced with the lepton number violation
for a natural realization of tiny neutrino masses.
 Such heavy Majorana neutrinos
can also be used for leptogenesis~\cite{Fukugita:1986hr}
to explain the baryon asymmetry of the Universe.

 There is an alternative scenario,
in which the smallness of neutrino masses
can be explained by the quantum effect.
 For example,
tiny Majorana masses
are radiatively generated at the one-loop level
in the Ma model~\cite{ref:Ma},
where new particles are not necessarily very heavy
and can be in the TeV-scale.
 Hence,
scenarios along this line can be in principle tested
directly by collider experiments.
 In addition,
new particles that are involved in the one-loop diagram
include the dark matter candidate.
 Similarly to the Ma model,
the Aoki-Kanemura-Seto~(AKS) model~\cite{ref:AKS}
gives tiny Majorana neutrino masses at the three-loop level,
where the dark matter candidate contributes
to the three-loop diagram.
 One of the remarkable features of this model
is that two $\SU(2)_L$-doublet Higgs fields are required.
 In general two Higgs doublet models can contain
CP violating phases in the Higgs sector.
 It is shown that
the strong first-order phase transition
is realized in the AKS model
which is required for successful electroweak baryogenesis~\cite{Kuzmin:1985mm}.
 Thus,
three problems can be simultaneously resolved
at the TeV-scale in the AKS model,
which can be probed at collider experiments.

 In Refs.~\cite{Kanemura:2015cca, Kanemura:2016ixx},
models for generating the neutrino mass matrix
are classified into several groups
by focusing on combinations of Yukawa matrices in the mass matrix.
 In the systematic study
for the Majorana neutrino mass generation~\cite{Kanemura:2015cca},
no combination other than the one in the AKS model
involves simultaneously the dark matter candidate
and the second $\SU(2)_L$-doublet Higgs field
that has the vacuum expectation value.
 On the other hand,
it was found in Ref.~\cite{Kanemura:2016ixx} that
masses of Dirac neutrinos can be radiatively generated
by using the dark matter candidate and
the second $\SU(2)_L$-doublet Higgs field%
~(See Figs.~11, 13 and 15 in Ref.~\cite{Kanemura:2016ixx}).

 In this letter,
we propose a concrete model,
where the combination of Yukawa matrices
to generate Dirac neutrino mass matrix
corresponds to the structure in Fig.~15 in Ref.~\cite{Kanemura:2016ixx}.
 As the lepton number violating phenomena
such as the neutrinoless double beta decay
have not been observed up to now,
it would be important to consider the possibility
that neutrinos are purely of the Dirac type fermions,
whose masses conserve the lepton number.
 Along this line,
we investigate the new model to simultaneously
provide the origin of tiny neutrino masses,
the dark matter candidate,
and the source of the baryon asymmetry of the Universe.
 We here ignore the CP violation in the scalar potential for simplicity
and concentrate on the realization of
the strong first-order phase transition,
which is necessary for successful electroweak baryogenesis.
 Calculation for the baryon asymmetry with the CP violation
is beyond the scope of this letter,
and leave it as our future work.

 This letter is organized as follows.
 In Sec.~II,
we present our model
where Dirac neutrino masses are generated
at the one-loop level.
 We see that our model
can be consistent with neutrino oscillation data
and the relic abundance of the dark matter
as well as the strong first-order
electroweak phase transition
required for the electroweak baryogenesis scenario.
 Section~III is devoted to further phenomenological studies
such as lepton flavor violating decays of charged leptons,
the spin-independent scattering cross section
of the dark matter on a proton,
and collider phenomenology.
 Conclusions are shown in Sec.~IV\@.
 Some formulae are presented in Appendix.



\section{The Model}
\label{sec:model}

\begin{table}
\begin{tabular}{c||
>{\centering\arraybackslash}p{10mm}|
>{\centering\arraybackslash}p{10mm}|
>{\centering\arraybackslash}p{10mm}|
>{\centering\arraybackslash}p{10mm}|
>{\centering\arraybackslash}p{10mm}||
>{\centering\arraybackslash}p{10mm}|
>{\centering\arraybackslash}p{10mm}|
>{\centering\arraybackslash}p{10mm}}
{}
 & \multicolumn{5}{c||}{ $Z_2$-even }
 & \multicolumn{3}{c}{ $Z_2$-odd }
\\
\cline{2-9}
{}
 & $\ell_R^{}$
 & $\nu_R^{}$
 & $\Phi_1$
 & $\Phi_2$
 & $s_3^+$
 & $\psi_R^0$
 & $s_2^0$
 & $s_2^+$
\\\hline\hline
Spin
 & $1/2$
 & $1/2$
 & $0$
 & $0$
 & $0$
 & $1/2$
 & $0$
 & $0$
\\\hline
$\SU(2)_L$
 & $\bf\underline{1}$
 & $\bf\underline{1}$
 & $\bf\underline{2}$
 & $\bf\underline{2}$
 & $\bf\underline{1}$
 & $\bf\underline{1}$
 & $\bf\underline{1}$
 & $\bf\underline{1}$
\\\hline
$\U(1)_Y$
 & $-1$
 & $0$
 & $1/2$
 & $1/2$
 & $1$
 & $0$
 & $0$
 & $1$
\\\hline
Lepton Number
 & $1$
 & $1$
 & $0$
 & $0$
 & $0$
 & $0$
 & $-1$
 & $-1$
\\\hline
$Z_2^\prime$
 & Even
 & Odd
 & Even
 & Even
 &
 & Even
 & Odd
 & Even
\\\hline
$Z_2^\dprime$
 & Odd
 & Even
 & Odd
 & Even
 & Odd
 & Even
 & Even
 & Odd
\end{tabular}
\caption{
 The list of new fields in our model,
which are added to the SM\@.
 A scalar field $s_3^+$
can be both of even and odd
for the $Z_2^\prime$ symmetry.
}
\label{tab:particle}
\end{table}

\subsection{ Particle contents and Lagrangian }

 In our model,
the lepton number~(\LN) conservation is imposed%
\footnote{
 The spharelon process breaks $\BN + \LN$ at the finite temperature,
where $\BN$ denotes the baryon number.
 The process is utilized for the electroweak baryogenesis scenario.
}.
 We introduce gauge singlet fermions
$\nu_{iR}^{}$~($i = 1 \text{--} 3$) with $\LN = 1$,
which result in right-handed components of Dirac neutrinos.
 New fields that are introduced to the SM
are listed in Table~\ref{tab:particle}.
 Fermions $\psi_{aR}^0$~($a = 1 \text{--} 3$)
are also gauge singlet
though they have $\LN = 0$
in comparison with $\LN = 1$ for $\nu_R^{}$.
 This model involves two $\SU(2)_L$-doublet scalar fields
$\Phi_1 \equiv (\phi_1^+, (v_1 + \phi_{1r}^0 + i \phi_{1i}^0)/\sqrt{2})^T$
and $\Phi_2 \equiv (\phi_2^+, (v_2 + \phi_{2r}^0 + i \phi_{2i}^0)/\sqrt{2})^T$
with $Y=1/2$ and $\LN = 0$,
where $v^2 = v_1^2 + v_2^2 = (246\,\GeV)^2$.
 Scalar fields $s_2^+$ and $s_3^+$ are $\SU(2)_L$-singlet with $Y = 1$;
the former has $\LN = -1$ while the latter does $\LN = 0$.
 The complex scalar field $s_2^0$ is a gauge singlet with $\LN = -1$.

 If the Dirac neutrino mass is simply generated
via the Yukawa interaction $y_\nu \overline{L} \wt{\Phi} \nu_R$,
where $\wt{\Phi} \equiv \epsilon \Phi^\ast$,
the size of the Yukawa coupling constant is extremely small%
~($y_\nu \sim 10^{-12}$),
which seems to be unnatural
and cannot be experimentally tested.
 In our model,
the Yukawa interaction is forbidden at the tree-level
by imposing a softly broken $Z_2$ symmetry
(we refer to it as the $Z_2^\prime$ symmetry),
under which $\nu_R$ is odd
while $\SU(2)_L$-doublet lepton and scalars are even.
 Properties of additional particles
with respect to the $Z_2^\prime$
are also shown in Table~\ref{tab:particle}.
 The scalar field $s_3^+$ can be both of even and odd.
 The $Z_2^\prime$ is softly broken in the scalar potential,
and then the Dirac neutrino mass can be generated
at the two-loop level as we see later.


 Relevant parts of Yukawa interactions
for generating neutrino masses are given by
\begin{eqnarray}
{\mathcal L}_\text{Yukawa}
=
 y_\ell \overline{L_\ell} \Phi_1 \ell_R
 + (Y_\psi^+)_{\ell a} \overline{(\ell_R)^c}\, \red{\psi_{a R}^0\, s_2^+}
 + (Y_\psi^0)_{i a} \overline{(\nu_{iR}^{})^c}\, \red{\psi_{a R}^0\, s_2^0}
 + \text{h.c.}
\label{eq:Yukawa}
\end{eqnarray}
 In order to forbid
the flavor changing neutral current~(FCNC) at the tree level,
we impose another softly-broken $Z_2$ symmetry%
~(we represent it as $Z_2^\dprime$)
as shown in Table.~\ref{tab:particle}.
 Quarks and the lepton doublet field are $Z_2^\dprime$-even.
 Then,
$\ell_R$ has the Yukawa interaction only with $\Phi_1$
similarly to the the type-X two Higgs doublet model~\cite{Aoki:2009ha}
(see also Refs.~\cite{Barger:1989fj, Grossman:1994jb, ref:THDM-Akeroyd}).



 The scalar potential is given by
\begin{eqnarray}
V
&=&
 V_\text{THDM}
 + m_{s0}^2 |\red{s_2^0}|^2
 + m_{s2}^2 |\red{s_2^+}|^2
 + m_{s3}^2 |s_3^+|^2
\nonumber\\
&&{}
 + \left(
    \mu_3
    \Bigl[
     \Phi_2^\dagger \epsilon \Phi_1^\ast s_3^+
    \Bigr] + \text{h.c.}
   \right)
 + \left(
    \mu_3^\prime
    \Bigl[
     s_3^- \red{s_2^+ s_2^{0\ast}}
    \Bigr] + \text{h.c.}
   \right)
\nonumber\\
&&{}
 + \lambda_{s0s2} |\red{s_2^0}|^2 |\red{s_2^+}|^2
 + \lambda_{s0s3} |\red{s_2^0}|^2 |s_3^+|^2
 + \lambda_{s2s3} |\red{s_2^+}|^2 |s_3^+|^2
\nonumber\\
&&{}
 + \lambda_{\phi 1 s0} \Phi_1^\dagger \Phi_1 |\red{s_2^0}|^2
 + \lambda_{\phi 1 s2} \Phi_1^\dagger \Phi_1 |\red{s_2^+}|^2
 + \lambda_{\phi 1 s3} \Phi_1^\dagger \Phi_1 |s_3^+|^2
\nonumber\\
&&{}
 + \lambda_{\phi 2 s0} \Phi_2^\dagger \Phi_2 |\red{s_2^0}|^2
 + \lambda_{\phi 2 s2} \Phi_2^\dagger \Phi_2 |\red{s_2^+}|^2
 + \lambda_{\phi 2 s3} \Phi_2^\dagger \Phi_2 |s_3^+|^2
\nonumber\\
&&{}
 + \lambda_{s0} |\red{s_2^0}|^4
 + \lambda_{s2} |\red{s_2^+}|^4
 + \lambda_{s3} |s_3^+|^4 ,
\end{eqnarray}
where $V_\text{THDM}$ is the following one
in two Higgs double models without the tree-level FCNC:
\begin{eqnarray}
V_\text{THDM}
&=&
 m_{11}^2 \Phi_1^\dagger \Phi_1
 + m_{22}^2 \Phi_2^\dagger \Phi_2
 - \left(
    m_{12}^2 \Phi_1^\dagger \Phi_2
    + \text{h.c.}
   \right)
\nonumber\\
&&\hspace*{0mm}{}
 + \frac{\lambda_1}{2} (\Phi_1^\dagger \Phi_1)^2
 + \frac{\lambda_2}{2} (\Phi_2^\dagger \Phi_2)^2
 + \lambda_3 (\Phi_1^\dagger \Phi_1) (\Phi_2^\dagger \Phi_2)
\nonumber\\
&&\hspace*{0mm}{}
 + \lambda_4 (\Phi_1^\dagger \Phi_2) (\Phi_2^\dagger \Phi_1)
 + \left(
    \frac{\lambda_5}{2}
    (\Phi_1^\dagger \Phi_2)^2
    + \text{h.c.}
   \right) .
\end{eqnarray}
 The complex phases of 
$\mu_3^{}$, $\mu_3^\prime$, and $\lambda_5$ can be eliminated
by the rephasing of $s_3^+$, $s_2^+$~(or $s^0$),
and $\Phi_1$~(or $\Phi_2$), respectively.
 For simplicity in this letter,
we take $m_{12}^2$ as a real parameter
and also assume that the CP is not spontaneously violated.
 $Z_2^\prime$ and $Z_2^\dprime$ symmetries are
softly broken by $\mu_3 \mu_3^\prime$ and $m_{12}^2$,
respectively.



 By virtue of the assignments of
conserved lepton numbers to new fields,
there appears an unbroken $Z_2$ symmetry
such that $\psi_R^0$, $s_2^0$, and $s_2^+$ have the odd parity.
 The $Z_2$ symmetry can be utilized
for the stabilization of the dark matter candidate.

 $Z_2$-odd scalar fields $s_2^0$ and $s_2^+$
are mass eigenstates ${\mathcal H}^0$ and ${\mathcal H}^+$
without mixings with other fields, respectively.
 Their masses are calculated as
\begin{eqnarray}
m_{{\mathcal H}^0}^2
&=&
 m_{s0}^2
 + \frac{ \lambda_{\phi 1 s0} v_1^2}{ 2 }
 + \frac{ \lambda_{\phi 2 s0} v_2^2}{ 2 } ,
\label{eq:mclH}
\\
%
%
m_{{\mathcal H}^+}^2
&=&
 m_{s2}^2
 + \frac{ \lambda_{\phi 1 s2} v_1^2}{ 2 }
 + \frac{ \lambda_{\phi 2 s2} v_2^2}{ 2 } .
\label{eq:mclHP}
\end{eqnarray}

 Three $Z_2$-even charged scalar fields
($\phi_1^+$, $\phi_2^+$, and $s_3^+$)
result in three mass eigenstates
($H_1^+$, $H_2^+$, and the Nambu-Goldstone boson $G^+$)
as
\begin{eqnarray}
\begin{pmatrix}
 G^+\\
 H_1^+\\
 H_2^+
\end{pmatrix}
=
 \begin{pmatrix}
  \cos\beta & \sin\beta & 0\\
  - \sin\beta \cos\theta_+ & \cos\beta \cos\theta_+ & \sin\theta_+\\
  \sin\beta \sin\theta_+ & - \cos\beta \sin\theta_+ & \cos\theta_+
 \end{pmatrix}
 \begin{pmatrix}
  \phi_1^+\\
  \phi_2^+\\
  s_3^+
 \end{pmatrix} ,
\end{eqnarray}
where $\tan\beta = v_2/v_1$.
 The mixing angle $\theta_+$ can be expressed as
\begin{eqnarray}
\tan(2\theta_+)
&=&
 \frac{ - 2 (M_{H^+}^{\prime 2})_{12} }
      {
        (M_{H^+}^{\prime 2})_{22} - (M_{H^+}^{\prime 2})_{11}
      } ,
\end{eqnarray}
where the $2\times 2$ matrix $M_{H^+}^{\prime 2}$ is defined as
\begin{eqnarray}
M_{H^+}^{\prime 2}
\equiv
 \begin{pmatrix}
  \displaystyle
  \frac{v^2}{v_1 v_2} m_{12}^2
  - \frac{1}{\,2\,} (\lambda_4 + \lambda_5) v^2
  &\displaystyle
    \frac{ v \mu_3 }{\sqrt{2}}
\\
  \displaystyle
  \frac{ v \mu_3 }{\sqrt{2}}
  &\displaystyle
    m_{s3}^2
    + \frac{ \lambda_{\phi1 s3} v_1^2 }{ 2 }
    + \frac{ \lambda_{\phi2 s3} v_2^2 }{ 2 }
 \end{pmatrix} .
\end{eqnarray}
 Masses of $H_1^+$ and $H_2^+$ are given by
\begin{eqnarray}
m_{H_1^+}^2
&=&
 \frac{1}{\,2\,}
 \left\{
  (M_{H^+}^{\prime 2})_{11}
  + (M_{H^+}^{\prime 2})_{22}
  - \sqrt{
          ( (M_{H^+}^{\prime 2})_{22} - (M_{H^+}^{\prime 2})_{11} )^2
          + 4 (M_{H^+}^{\prime 2})_{12}^2
         }
 \right\} ,
\\
%
%
m_{H_2^+}^2
&=&
 \frac{1}{\,2\,}
 \left\{
  (M_{H^+}^{\prime 2})_{11}
  + (M_{H^+}^{\prime 2})_{22}
  + \sqrt{
          ( (M_{H^+}^{\prime 2})_{22} - (M_{H^+}^{\prime 2})_{11} )^2
          + 4 (M_{H^+}^{\prime 2})_{12}^2
         }
 \right\} .
\end{eqnarray}

 $Z_2$-even neutral scalars%
~(CP-even $h$ and $H$, CP-odd $A$,
and the Nambu-Goldstone boson $G^0$)
are constructed from $\Phi_1$ and $\Phi_2$
similarly to two Higgs doublet models,
where $h$ is the discovered Higgs boson
with the mass $m_h = 125\,\GeV$.
 Throughout this letter,
we take $\sin(\beta - \alpha) = 1$,
where $\alpha$ denotes the mixing angle
between $\phi_{1r}^0$ and $\phi_{2r}^0$.


 Differences of our model from the AKS model are as follows.

\noindent
i) Neutrinos are Dirac fermions,
for which right-handed neutrinos $\nu_R^{}$ are introduced
with the lepton number conservation,
while the AKS model is for Majorana neutrino masses
without $\nu_R^{}$.

\noindent
ii) Assignments of conserving lepton numbers
to new fields stabilize the dark matter candidate
instead of the imposed $Z_2$ symmetry in the AKS model.

\noindent
iii) The $Z_2$-odd neutral scalar $s_2^0$
is a complex field with the conserved lepton number
while the AKS model involves a $Z_2$-odd real scalar field.

\noindent
iv) The $Z_2$-even $\SU(2)_L$-singlet charged scalar $s_3^+$
is introduced, which is absent in the AKS model.

\begin{figure}[t]
\includegraphics[scale=0.7]{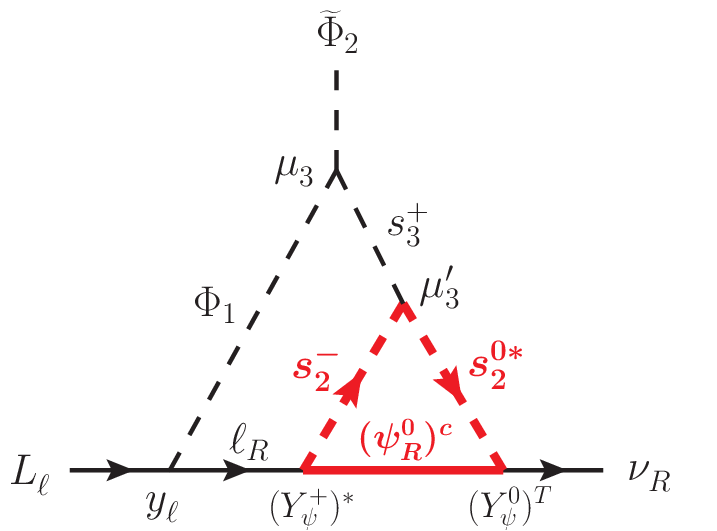}
\caption{
 The diagram for generating Dirac neutrino masses in our model.
 The arrow shows the flow of the lepton number,
which is conserved in our model.
}
\label{fig:mD}
\end{figure}

\subsection{ Dirac Neutrino Masses }

 Neutrinos acquire Dirac masses at the two-loop level
as $(m_\text{D})_{\ell i}^{}\, \overline{\nu_{\ell L}^{}}\, \nu_{iR}^{}$
via the diagram in Fig.~\ref{fig:mD}.
 Notice that
the interaction $\Phi_2^\dagger \epsilon \Phi_1^\ast s_2^+ s_2^0$
is forbidden because this is a hard breaking term of $Z_2^\prime$.
 The coupling constant $\mu_3$ can be expressed by $\theta_+$.
 Although arbitrary number of $|\Phi_i|^2$ can be attached
to each of scalar lines in the diagram with
appropriate coupling constant $\lambda$,
such effects to neutrino masses
are involved in scalar mass eigenvalues
as $\lambda v_i^2$.
 As a result,
the mass matrix $(m_\text{D})_{\ell i}$ is given by
\begin{eqnarray}
(m_\text{D}^{})_{\ell i}^{}
&=&
 \frac{1}{\sqrt{2}\, v}\,
 m_\ell^{}
 \left(
  m_{H_2^+}^2 - m_{H_1^+}^2
 \right)
 \mu_3^\prime
 \tan\beta\,
 \sin(2\theta_+)
 \sum_a
 (Y_\psi^+)^\ast_{\ell a} (Y_\psi^0)^T_{a i} I_{\ell a},
\label{eq:mD}
\end{eqnarray}
where the explicit formula for the two-loop function $I_{\ell a}$
is presented in Appendix~\ref{sec:app-A}.
 We can take the basis where
$\nu_{iR}^{}$ are already mass eigenstates
without loss of generality.
 Then,
the mass matrix can be expressed as
$m_\text{D}^{} = U_\MNS\,\text{diag}(m_1, m_2, m_3)$
with neutrino mass eigenvalues $m_i$~($i = 1 \text{-} 3$).
 The Maki-Nakagawa-Sakata matrix~\cite{Maki:1962mu} $U_\MNS$
can be parametrized as
\begin{eqnarray}
U_\MNS
=
 \begin{pmatrix}
  1 & 0 & 0\\
  0 & c_{23} & s_{23}\\
  0 & -s_{23} & c_{23}
 \end{pmatrix}
 \begin{pmatrix}
  c_{13} & 0 & s_{13} e^{-i \delta}\\
  0 & 1 & 0\\
  -s_{13} e^{i\delta} & 0 & c_{13}
 \end{pmatrix}
 \begin{pmatrix}
  c_{12} & s_{12} & 0\\
  -s_{12} & c_{12} & 0\\
  0 & 0 & 1
 \end{pmatrix} ,
\end{eqnarray}
where $c_{ij} \equiv \cos\theta_{ij}$
and $s_{ij} \equiv \sin\theta_{ij}$.
 Neutrino oscillation data result in
 We take the following values
consistent with neutrino oscillation data:
\begin{eqnarray}
&&
\sin^2\theta_{23} = 0.514~\text{\cite{Abe:2015awa}} , \quad
\sin^2(2\theta_{13}) = 0.0841~\text{\cite{An:2016ses}} , \quad
\tan^2\theta_{12} = 0.427~\text{\cite{Aharmim:2011vm}} ,
\\
%
%
&&
\Delta m^2_{32} = 2.51\times 10^{-3}\,\eV^2~\text{\cite{Abe:2015awa}} , \quad
\Delta m^2_{21} = 7.46\times 10^{-5}\,\eV^2~\text{\cite{Aharmim:2011vm}} ,
\end{eqnarray}
where $\Delta m^2_{ij} \equiv m_i^2 - m_j^2$.
 For example,
$m_\text{D}$ for these values
with $m_3 > m_1 = 0.01\,\eV$ and $\delta = 0$
can be generated in our model by the following
benchmark set of parameter values:
\begin{equation}
\begin{split}
&
Y_\psi^+
=
 \begin{pmatrix}
  1 & 0.01 & 0.01\\
  0.01 & 1 & 0.01\\
  0.01 & 0.01 & 1
 \end{pmatrix} ,\quad
%
%
Y_\psi^0
\simeq
 \begin{pmatrix}
  3.2\times 10^{-1}
   & -4.0\times 10^{-3}
   & -3.1\times 10^{-3}\\
  2.7\times 10^{-1}
   & -1.4\times 10^{-3}
   & -2.8\times 10^{-3}\\
  2.9\times 10^{-1}
   & 3.9\times 10^{-3}
   & -2.6\times 10^{-3}
 \end{pmatrix} ,
\\
%
%
&
\mu_3^\prime = 50\,\GeV, \quad
\tan\beta = 3, \quad
\sin\theta_+ = 0.1 ,
\\
%
%
&
m_{H_1^+}^{} = 200\,\GeV, \quad
m_{H_2^+}^{} = 300\,\GeV, \quad
m_{{\mathcal H}^+}^{} = 350\,\GeV, \quad
m_{{\mathcal H}^0}^{} = 63\,\GeV ,
\\
%
%
&
m_{\psi_{1R}}^{} = m_{\psi_{2R}}^{} = m_{\psi_{3R}}^{}
= 5\,\TeV .
\end{split}
\label{eq:benchmark}
\end{equation}
 We see that
tiny Dirac neutrino masses are obtained
without extremely small values of parameters
and very heavy particles.
 Notice that ${\mathcal H}^0$ for the benchmark set
is the lightest $Z_2$-odd particle
and the dark matter candidate.

 For different values of parameters,
one of $\psi_{iR}^0$ can be the lightest $Z_2$-odd particle
and the dark matter candidate.


\subsection{ Relic Abundance of the Dark Matter }

\begin{figure}[t]
 \begin{minipage}[c]{50mm}
  \includegraphics[scale=0.6]{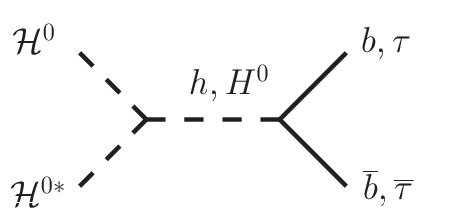}
 \end{minipage}
 \begin{minipage}[c]{50mm}
  \includegraphics[scale=0.6]{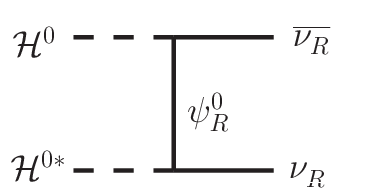}
 \end{minipage}
 \begin{minipage}[c]{50mm}
  \includegraphics[scale=0.6]{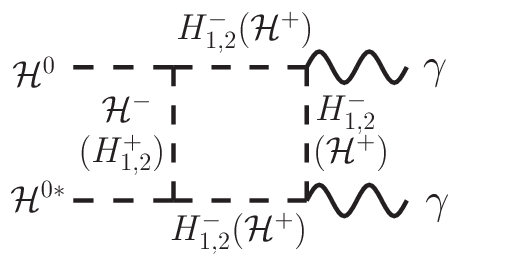}
 \end{minipage}\\[1mm]
 \begin{minipage}[c]{50mm}
  (a)
 \end{minipage}
 \begin{minipage}[c]{50mm}
  (b)
 \end{minipage}
 \begin{minipage}[c]{50mm}
  (c)
 \end{minipage}\\[4mm]
 \begin{minipage}[c]{50mm}
  \includegraphics[scale=0.6]{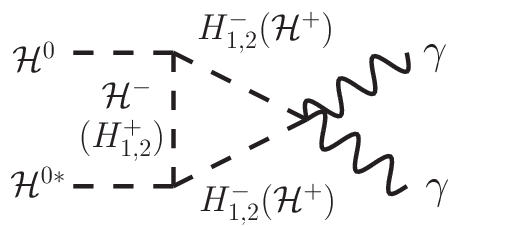}
 \end{minipage}
 \begin{minipage}[c]{50mm}
  \includegraphics[scale=0.6]{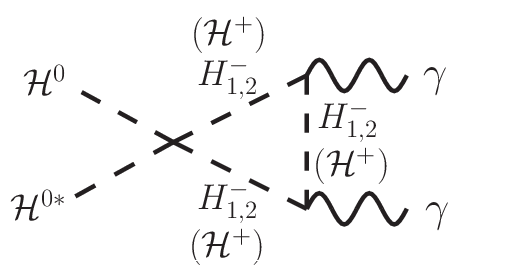}
 \end{minipage}
 \begin{minipage}[c]{50mm}
  \includegraphics[scale=0.6]{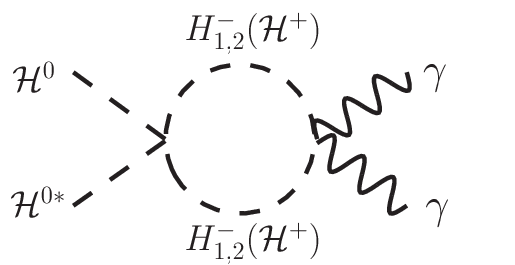}
 \end{minipage}\\[1mm]
 \begin{minipage}[c]{50mm}
  (d)
 \end{minipage}
 \begin{minipage}[c]{50mm}
  (e)
 \end{minipage}
 \begin{minipage}[c]{50mm}
  (f)
 \end{minipage}\\[4mm]
 \begin{minipage}[c]{50mm}
  \includegraphics[scale=0.6]{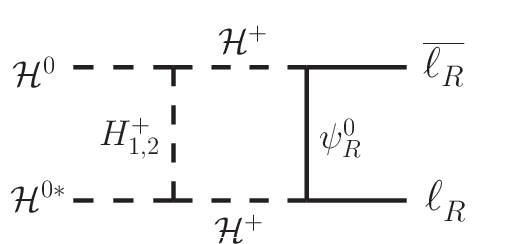}
 \end{minipage}
 \begin{minipage}[c]{50mm}
  \includegraphics[scale=0.6]{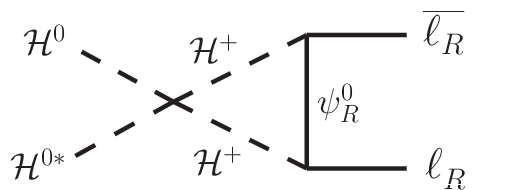}
 \end{minipage}\\[1mm]
 \begin{minipage}[c]{50mm}
  (g)
 \end{minipage}
 \begin{minipage}[c]{50mm}
  (h)
 \end{minipage}\\[-2mm]
\caption
{
 Diagrams of the pair-annihilation of ${\mathcal H}^0$.
}
\label{fig:pair}
\end{figure}

\begin{figure}[t]
\includegraphics[scale=0.6]{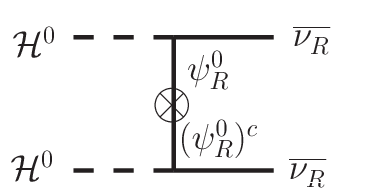}\\[-4mm]
\caption
{
 The diagram of the self-annihilation of ${\mathcal H}^0$.
}
\label{fig:self}
\end{figure}

\begin{figure}[t]
\includegraphics[scale=0.9]{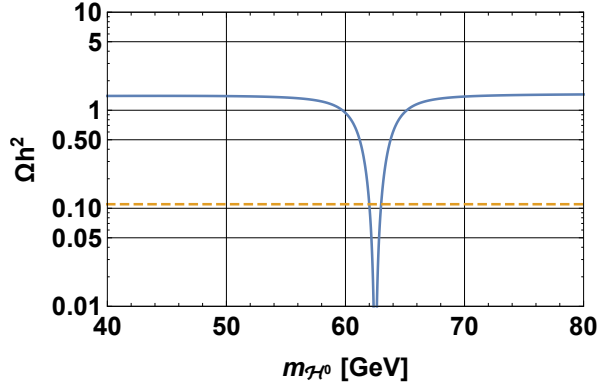}
\caption
{
 The relic abundance of the scalar dark matter ${\mathcal H}^0$ in our model.
 The horizontal dashed line corresponds to
$\Omega h^2 = 0.1186$~\cite{Ade:2015xua}.
}
\label{fig:abundance}
\end{figure}

 For the benchmark set in eq.~\eqref{eq:benchmark},
the dark matter candidate~(the lightest $Z_2$-odd particle) is
a complex scalar particle ${\mathcal H}^0$.
 Its relic abundance%
~($\Omega_{{\mathcal H}^0} h^2 + \Omega_{{\mathcal H}^{0\ast}} h^2
\simeq 2 \Omega_{{\mathcal H}^0} h^2$)
must be consistent with
the experimental constraint
$\Omega_\DM h^2 = 0.1186 \pm 0.0020$~\cite{Ade:2015xua}.
 Figures~\ref{fig:pair} and \ref{fig:self}
show diagrams of pair and self-annihilation of ${\mathcal H}^0$,
respectively%
\footnote{
 Contribution of
${\mathcal H}^0 {\mathcal H}^{0\ast} \to h(H) \to \gamma\gamma$
is always subdominant
while loop diagrams for
${\mathcal H}^0 {\mathcal H}^{0\ast} \to \gamma\gamma$
in Fig.~\ref{fig:pair}
can be dominant ones in some parameter region.
}.
 See Appendix~\ref{sec:app-B}
for formulae of cross sections for them.
 The cross section $\sigma$,
which is the sum of cross sections for these annihilation processes,
can be expanded as $\sigma v_\rel^{} \simeq \sigma_0 + \sigma_1 v_\rel^2$
with respect to the relative velocity $v_\rel^{}$ between initial particles.
 Then,
the relic abundance of ${{\mathcal H}^0}$
can be calculated~\cite{Kolb:1990vq} with
\begin{eqnarray}
\Omega_{{\mathcal H}^0} h^2
&=&
 1.04 \times 10^9
 \frac{ \sqrt{g_\ast} }{ g_{\ast S} }
 \frac{ \GeV }{ m_\Pl }
 \frac{ \GeV^{-2} }{ \sigma_0 }
 x_f
 \left(
  1 + \frac{ 3 \sigma_1 }{ \sigma_0 } x_f^{-1}
 \right)^{-1} ,
\\
%
%
x_f
&=&
 \ln\Bigl[
     0.038\, \frac{g_{{\mathcal H}^0}^{}}{\sqrt{g_\ast}}\,
     m_\Pl^{}\, m_{{\mathcal H}^0}^{}\, \sigma_0
    \Bigr]
 -
 \frac{1}{\,2\,}
 \ln\Bigl[
     \Bigl\{
      0.038\, \frac{g_{{\mathcal H}^0}^{}}{\sqrt{g_\ast}}\,
      m_\Pl^{}\, m_{{\mathcal H}^0}^{}\, \sigma_0
     \Bigr\}
    \Bigr]
\nonumber\\*
&&\hspace*{30mm}{}
 +
 \ln\Bigl[
     1
     +
     \frac{ 6 \sigma_1 }{ \sigma_0 }
       \Bigl\{
        \ln\Bigl(
            0.038\, \frac{g_{{\mathcal H}^0}^{}}{\sqrt{g_\ast}}\,
            m_\Pl^{}\, m_{{\mathcal H}^0}^{}\, \sigma_0
           \Bigr)
       \Bigr\}^{-1}
    \Bigr] ,
\end{eqnarray}
where $m_\Pl^{} = 1.2 \times 10^{19}\,\GeV$ is the Planck mass,
$g_\ast = g_{S\ast} = 106.75$,
and $g_{{\mathcal H}^0}^{} = 1$.
 Figure~\ref{fig:abundance} is obtained
by using the benchmark set in eq.~\eqref{eq:benchmark}%
~($m_{{\mathcal H}^0}^{}$ is taken as the $x$-axis) with
\begin{eqnarray}
 m_H^{} = 200\,\GeV, \quad
 \lambda_{\phi1 s0} = 0.02, \quad
 \lambda_{\phi2 s0} = 0.005 .
\label{eq:benchmark-2}
\end{eqnarray}
 We see that
the observed value of $\Omega_\DM h^2$%
~(horizontal dashed line in Fig.~\ref{fig:abundance})
is produced for $m_{{\mathcal H}^0}^{} = 63\,\GeV$
of the benchmark set in eq.~\eqref{eq:benchmark}.
 Since $m_{{\mathcal H}^0}^{} \simeq m_h^{}/2$ for the benchmark set,
the $h$-mediation is
the dominant annihilation process.
 Notice that
the cross section for the self-annihilation in Fig.~\ref{fig:self}
is almost independent of $m_{{\mathcal H}^0}^{}$
if $m_{{\mathcal H}^0}^2 \ll m_\psi^2$.
 Thus,
even for a different value of $m_{{\mathcal H}^0}^{}$,
the constraint on $\Omega_\DM h^2$ can be easily satisfied
by adjusting the overall scale of $Y_\psi^0$,
which does not affect $\ell \to \ell^\prime \gamma$
and $\mu \to \overline{e} ee$;
 the neutrino mass scale can be recovered
to the appropriate value
by changing $\mu^\prime$ for example.


\subsection{ Strong First-Order Phase Transition }

\begin{figure}[t]
\includegraphics[scale=0.46]{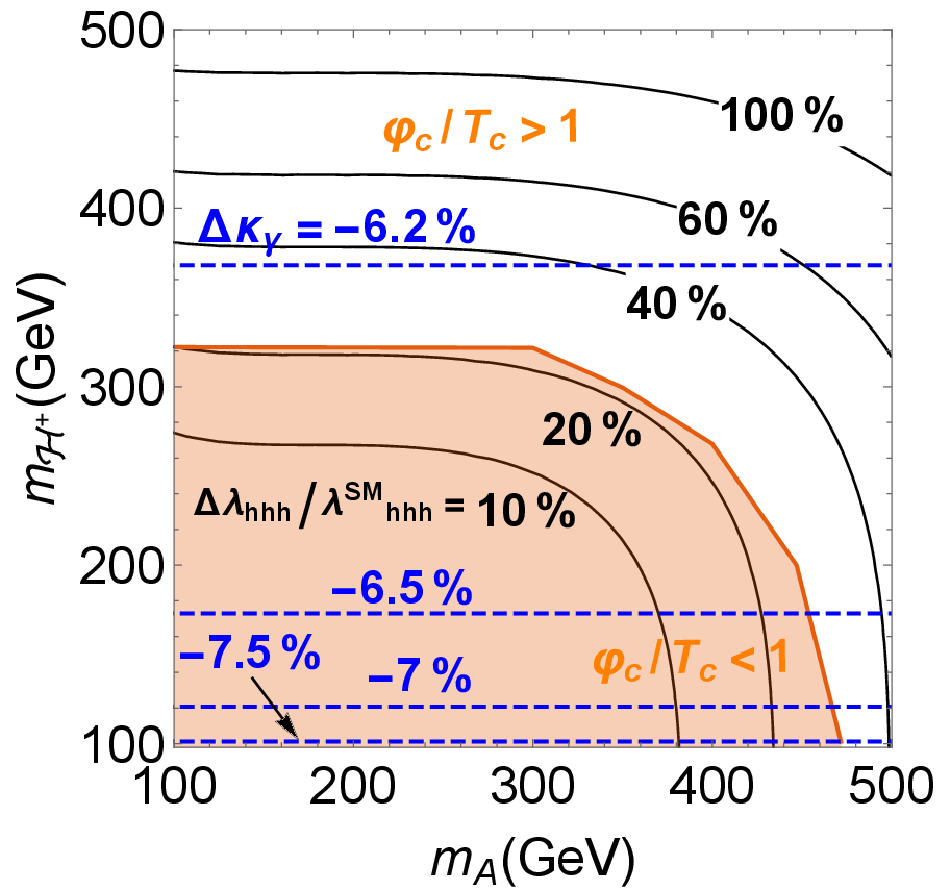}
\caption
{
 In the white region,
our model satisfies
the condition $\varphi_c/T_c \gtrsim 1$
for the strong first-order phase transition,
which is required for the electroweak baryogenesis scenario.
 Solid lines are contours of
the deviation $\Delta \lambda_{hhh}/\lambda_{hhh}^\text{SM}$
of $\lambda_{hhh}$ from the SM value $\lambda_{hhh}^\text{SM}$.
 Dashed lines show contours of the deviation $\Delta \kappa_\gamma$
of $\Gamma(h \to \gamma\gamma)$ from the SM value.
}
\label{fig:phic_Tc}
\end{figure}

\begin{figure}[t]
\includegraphics[scale=0.5]{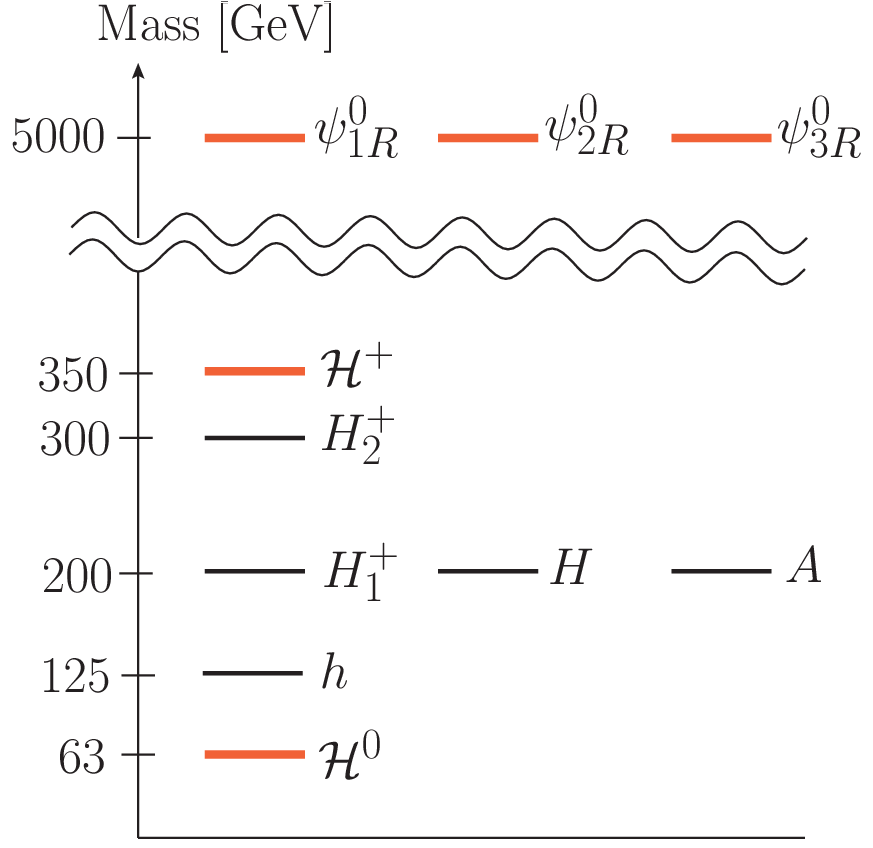}
\caption
{
 Mass spectrum of $h$ and new particles
for the benchmark values
in eqs.~\eqref{eq:benchmark} and \eqref{eq:benchmark-2}
with $m_A = 200\,\GeV$.
 Red bold bars show masses of $Z_2$-odd particles.
}
\label{fig:spectrum}
\end{figure}

 Baryon asymmetry of the Universe can be generated
if the electroweak symmetry is broken
via the strong first-order phase transion.
 Let us examine below
whether the strong first-order phase transion
can be achieved in our model or not.
 The effective potential $V_\eff$ at the one-loop level
at the finite temperature $T$ is given by
the sum of the potential $V_\tree$ at the tree level,
the one-loop correction $\Delta V$
at the zero temperature~\cite{DiLuzio:2014bua},
the correction $\Delta V_T$
at the finite temperature~\cite{Dolan:1973qd}
as follows:
\begin{eqnarray}
V_\eff(\varphi, T)
&=&
 V_\tree(\varphi)
 + \Delta V(\varphi)
 + \Delta V_T(\varphi, T)
 + \Delta V_\text{ring},
\\
%
%
V_\tree(\varphi)
&=&
 - \frac{m_h^2}{4} \varphi^2
 + \frac{m_h^2}{8v^2} \varphi^4
 + \frac{1}{\,2\,} A \varphi^2 ,
\\
%
%
\Delta V(\varphi)
&=&
 \frac{1}{64\pi^2}
 \sum_i
  n_i (-1)^{2 s_i} \wt{m}_i^4(\varphi)
  \left(
   \ln\frac{ \wt{m}_i^2(\varphi) }{ Q^2 } - c_i
  \right) ,
\\
%
%
\Delta V_T(\varphi, T)
&=&
 \frac{T^4}{2\pi^2}
 \sum_i
  n_i I_i\!\left( \frac{ \wt{m}_i^2(\varphi) }{T^2} \right) ,
\end{eqnarray}
where $\varphi$ is the order parameter
for the electroweak symmetry breaking,
$n_i$ denotes the degree of freedom
for the particle $i$~($= t, W^\pm, Z, \gamma$, and scalar bosons),
$s_i$ is the spin.
 The field-dependent masses
$\widetilde{m}_i^{}(\varphi)$
are shown in Appendix~\ref{sec:app-C}.
 Contributions from the ring diagrams~\cite{Carrington:1991hz}
are taken into account by replacing
$\widetilde{m}_i^{}(\varphi)$
with $\widetilde{m}_i^{}(\varphi, T)$
at a finite temperature $T$,
which are given in Appendix~\ref{sec:app-D}.
 The constant $c_i$ is $3/2$ for fermions and scalar bosons
while it is $5/6$ for gauge bosons.
 The counter term $A$ and the renormalization scale $Q$
can be determined by renormalization conditions
$\partial V_\eff(\varphi, 0)/\partial \varphi|_{\varphi=v} = 0$
and $\partial^2 V_\eff(\varphi, 0)/\partial \varphi^2|_{\varphi=v} = m_h^2$,
where the infrared diverging $\ln \wt{m}_i^2(v)$
of the Nambu-Goldstone bosons~($i = G^0, G^\pm$)
are replaced with
$\text{Re} \int_0^1 dx \ln\{ -x(1-x) m_h^2 \}$~\cite{Cline:1996mga}.
 The distribution functions are given by
$I_i(\alpha^2)
= \int_0^\infty\!\! dx\,
  x^2 \ln\{ 1 \mp \exp(-\sqrt{x^2 + \alpha^2}) \}$,
where the minus sign is for bosons
and the plus sign is for fermions~\cite{Dolan:1973qd}.
 The critical temperature $T_c$
and the critical value $\varphi_c$~($\varphi$ at $T_c$)
are obtained with $V_\eff(\varphi_c, T_c) = 0$
and $\partial V_\eff(\varphi, T_c)/\partial \varphi|_{\varphi=\varphi_c} = 0$.
 Then,
$\varphi_c/T_c \gtrsim 1$ is required
for the strong first-order phase transition~\cite{Kuzmin:1985mm}.
 It is shown in Fig.~\ref{fig:phic_Tc} that
the condition is satisfied
in the white region%
~($m_A^{} \gtrsim 470\,\GeV$
and $m_{{\mathcal H}^\pm}^{} \gtrsim 300\,\GeV$).
 By taking $m_A = 200\,\GeV$ as a benchmark value
together with values in eqs.~\eqref{eq:benchmark} and \eqref{eq:benchmark-2},
the mass spectrum of scalar bosons and $\psi_R^0$ are
presented in Fig.~\ref{fig:spectrum}.




\section{Phenomenology}
\label{sec:pheno}


 The Yukawa interaction with $Y_\psi^+$ causes $\ell \to \ell^\prime \gamma$
via the ${\mathcal H}^+$-$\psi_R^0$ loop.
The branching ratio is calculated as
\begin{eqnarray}
\BR(\ell \to \ellp \gamma)
=
 \frac{ 3 \alpha_\EM^{} }
      { 64\pi G_F^2 m_{{\mathcal H}^+}^4 }
 \Bigl|
  (Y_\psi^+ (Y_\psi^+)^\dagger)_{\ell \ellp}
  F_2\Bigl( \frac{ m_\psi^2 }{ m_{{\mathcal H}^+}^2 } \Bigr)
 \Bigr|^2
 \BR(\ell \to e \nu \overline{\nu}) ,
\end{eqnarray}
where
$\alpha_\EM^{}$ denotes the fine structure constant,
$G_F$ is the Fermi coupling constant,
and
$
F_2(x)
\equiv
 ( 1 - 6x + 3x^2 + 2x^3 -6x^2\ln x )/\{ 6(1-x)^4 \}
$.
 The benchmark set in eq.~\eqref{eq:benchmark} result in
$\BR(\mu\to e\gamma) = 5.2\times 10^{-14}$,
$\BR(\tau \to e\gamma) = 8.8\times 10^{-15}$
and $\BR(\tau \to \mu\gamma) = 8.8\times 10^{-15}$,
which satisfy current experimental bounds
$\BR( \mu \to e \gamma ) < 4.2\times 10^{-13}$~\cite{TheMEG:2016wtm},
$\BR( \tau \to e \gamma ) < 3.3 \times 10^{-8}$~\cite{Aubert:2009ag}
and $\BR( \tau \to \mu \gamma ) < 4.4 \times 10^{-8}$~\cite{Aubert:2009ag}.
 The benchmark value for $\mu \to e\gamma$ is
close to the expected sensitivity
$\BR(\mu\to e\gamma) = 6\times 10^{-14}$
at the future MEG-II experiment~\cite{Baldini:2013ke}
while benchmark values for $\tau \to \ell\gamma$
are too small to be measured.


 The Yukawa interaction with $Y_\psi^+$
causes also $\mu \to \overline{e} ee$.
 Ignoring contributions of penguin diagrams
because of constraints for $\ell \to \ell^\prime \gamma$,
the branching ratio for $\mu \to \overline{e} ee$ via box diagrams
is given by
\begin{eqnarray}
&&
\BR( \mu \to \overline{e} ee )
=
 \frac{1}{4G_F^2}
 \biggl|
  \sum_{a,b}
   (Y_\psi^+)_{ea}^\ast
   (Y_\psi^+)_{ae}^T\,
   (Y_\psi^+)_{eb}^\ast
   (Y_\psi^+)_{b\mu}^T\,
   (I^{\text{box}}_1)_{ba}
\nonumber\\*
&&\hspace*{40mm}{}
  +
  \frac{1}{\,2\,}
  \sum_{a,b}
   (Y_\psi^+)_{e a}^\ast
   (Y_\psi^+)_{a e}^\dagger
   (Y_\psi^+)_{\mu b}
   (Y_\psi^+)_{b e}^T\,
   m_{\psi_b}^{} m_{\psi_a}^{}
   (I^{\text{box}}_2)_{ba}
 \biggr|^2 ,
\end{eqnarray}
where formulae of
loop functions $(I^{\text{box}}_1)_{ba}$ and $(I^{\text{box}}_2)_{ba}$
are shown in Appendix~\ref{sec:app-A}.
 The experimental constraint
$\BR( \mu \to \overline{e} ee ) < 1.0\times 10^{-12}$%
~\cite{Bellgardt:1987du}
is satisfied by the benchmark set in eq.~\eqref{eq:benchmark},
which gives $\BR( \mu \to \overline{e} ee ) = 1.0 \times 10^{-13}$.
 The benchmark value
can be measured at the future Mu3e experiment,
whose expected sensitivity is
$\BR( \mu \to \overline{e} ee ) \sim 10^{-16}$~\cite{Blondel:2013ia}.

 Although coupling constants $G_{\ell\ellp}^{}$
for decays $\ell \to \ellp \nu \overline{\nu}$
are universal in the SM
such that $G_{\ell\ellp}^{} = G_F^{}$,
they can be different from each other
due to contributions of new particles.
 The prediction
$G_{\tau e}^2/G_F^2 \simeq G_{\tau \mu}^2/G_F^2 \lesssim 1$
is obtained for the Group-V in Ref.~\cite{Kanemura:2016ixx}
by concentrating on the matrix structure
of the neutrino mass matrix.
 In our explicit model,
which belongs to the Group-V in Ref.~\cite{Kanemura:2016ixx},
we can calculate the size of the lepton universality violation.
 The benchmark set in eq.~\eqref{eq:benchmark}
results in
$G_{\tau e}^2/G_F^2 - 1 \sim G_{\tau \mu}^2/G_F^2 - 1 \sim -1 \times 10^{-15}$.
 These values are consistent with experimental bounds,
$G_{\tau e}^2/G_F^2 = 1.0029\pm 0.0046$~\cite{Olive:2016xmw},
$G_{\tau\mu}^2/G_F^2 = 0.981 \pm 0.018$~\cite{Olive:2016xmw},
and $G_{\tau\mu}^2/G_{\tau e}^2 = 1.0036 \pm 0.0020$~\cite{Aubert:2009qj}.
 Deviations of these benchmark values from unity
seem to be too small so that they cannot be measured.


 The spin-independent scattering cross section $\sigma_\SI^{}$
of the dark matter ${\mathcal H}^0$ on a proton can be calculated with
the following formulae:
\begin{eqnarray}
&&
\sigma_\SI^{}
=
 \frac{ m_p^2 }{ 4 \pi ( m_{{\mathcal H}^0}^{} + m_p )^2 }\,
 f_p^2 ,\\
%
%
&&
\frac{f_p}{m_p}
=
 \sum_{q = u,d,s} f_{Tq}^{(p)} \frac{ f_q }{ m_q }
 +
 \frac{2}{27} f_{TG}^{(p)} 
 \sum_{q = c,b,t} \frac{ f_q }{ m_q } , \qquad
%
%
f_{TG}^{(p)}
=
 1 - \sum_{q = u,d,s} f_{Tq}^{(p)} .
\end{eqnarray}
 The coupling constant $f_q$ of
the effective interaction
$f_q {\mathcal H}^0 {\mathcal H}^{0\ast} \overline{q} q$
in our model
is given by
\begin{eqnarray}
&&
\frac{ f_q }{ m_q }
=
 \frac{
       \lambda_{\phi 1 s0}\, \sin\alpha \cos\beta
       - \lambda_{\phi 2 s0}\, \cos\alpha \sin\beta
      }
      { m_h^2 }\,
 \frac{ \cos\alpha }{ \sin\beta }
\nonumber\\*
&&\hspace*{30mm}{}
 -
  \frac{
        \lambda_{\phi 1 s0}\, \cos\alpha \cos\beta
        + \lambda_{\phi 2 s0}\, \sin\alpha \sin\beta
       }
       { m_H^2 }\,
  \frac{ \sin\alpha }{ \sin\beta } ,
\label{eq:fq}
\end{eqnarray}
where $\alpha$ is the mixing angle
for $\phi_{1r}^0$ and $\phi_{2r}^0$.
 We take
$\sin(\beta - \alpha) = 1$,
$f_{Tu}^{(p)} = 0.016$ and $f_{Td}^{(p)} = 0.011$~\cite{Ellis:2008hf}
for $f_{Ts}^{(p)} = 0$ allowed by a lattice QCD calculation~\cite{ref:y=0}
in addition to values in eqs.~\eqref{eq:benchmark} and \eqref{eq:benchmark-2}.
 Then, we obtain
$\sigma_\SI^{} = 5.0\times 10^{-47}\,\text{cm}^2$,
which satisfies the current bound~%
($\sigma_\SI^{} \lesssim 10^{-46}\,\text{cm}^2$
for $m_\DM^{} \simeq 100\,\GeV$)~\cite{Akerib:2016vxi}.
 The benchmark value is in expected sensitivity regions of
the LZ experiment~\cite{Akerib:2015cja}
and the XENON1T experiment~\cite{Aprile:2015uzo}.


 There are three charged scalar particles in this model,
which can contribute to $h \to \gamma\gamma$.
 Contours of $\Delta \kappa_\gamma^{}$~($\equiv \kappa_\gamma^{} - 1$),
where
$\kappa_\gamma
\equiv
( \Gamma( h \to \gamma\gamma )/\Gamma( h \to \gamma\gamma )_\text{SM} )^{1/2}$,
are presented in Fig.~\ref{fig:phic_Tc} with dashed lines.
 The combined analysis of data
obtained in the ATLAS and the CMS collaborations
shows $|\kappa_\gamma^{}| = 0.90_{-0.09}^{+0.10}$~\cite{Khachatryan:2016vau}.
 The expected precision is $2$-$5\,\%$
when the $3000\,\fb^{-1}$ integrated luminosity is accumulated
at the LHC~\cite{Dawson:2013bba}.
 The precision at the ILC can be about $3\,\%$~\cite{Fujii:2015jha}
if $2000\,\fb^{-1}$ data with $\sqrt{s}=250\,\GeV$,
$200\,\fb^{-1}$ data with $\sqrt{s}=350\,\GeV$,
and $4000\,\fb^{-1}$ data with $\sqrt{s}=500\,\GeV$
are combined~(the H-20 operating scenario~\cite{Barklow:2015tja}).
 New charged scalars may also contribute to $h \to \gamma Z$.
 In our model,
negative deviations are predicted for the $h\gamma\gamma$ coupling
in most of the parameter space.
 The deviations are expected to be detected
in above the experiments.


 The $hhh$ coupling constant $\lambda_{hhh}$
can be evaluated by using $V_\eff$ at the zero temperature as
$\lambda_{hhh}^{}
= \partial^3 V_\eff(\varphi, 0)/\partial \varphi^3 \bigr|_{\varphi = v}$,
where diverging contributions of the Nambu-Goldstone bosons are removed.
 The deviation $\Delta \lambda_{hhh}$%
~($\equiv \lambda_{hhh} - \lambda_{hhh}^\text{SM}$)
of $\lambda_{hhh}$ in our model~($\theta_+ = 0$ for simplicity)
from the SM value $\lambda_{hhh}^\text{SM}$
is shown in Fig.~\ref{fig:phic_Tc}.
 For this benchmark point, we see that
$\Delta \lambda_{hhh}/\lambda_{hhh}^\text{SM} \gtrsim 20\,\%$
indicates $\varphi_c/T_c \gtrsim 1$.
 The high-luminosity LHC with $3000\,\fb^{-1}$
is expected to measure signal from
the Higgs boson pair production
with the $54\,\%$ uncertainty~\cite{CMS:2015nat}.
 Precision of the measurement of $\lambda_{hhh}$ at the ILC 
can be $26\,\%$ at $\sqrt{s} = 500\,\GeV$
for the H-20 operating scenario
and $10\,\%$ at $\sqrt{s} = 1\,\TeV$
with the $8000\,\fb^{-1}$ data~\cite{Fujii:2017ekh}.
 The electroweak baryogenesis scenario in our model
can be tested by precise measurements of $\lambda_{hhh}$
at these future experiments
similarly to the case in a two Higgs doublet model~\cite{Kanemura:2004ch}.


 The strongly first-order phase transition in the early Universe
can also be tested by detecting
the characteristic spectrum of gravitational waves~(GWs)%
~\cite{Apreda:2001us, Grojean:2006bp, Espinosa:2008kw, Caprini:2015zlo,
Huber:2015znp, Chala:2016ykx, Kobakhidze:2016mch}
at future space based GW interferometers
such as LISA~\cite{Audley:2017drz},
DECIGO~\cite{Kawamura:2011zz} and BBO~\cite{Corbin:2005ny}.
 In particular,
it has been decided that
LISA will start in relatively near future~\cite{Audley:2017drz}.
 In Refs.~\cite{Kakizaki:2015wua, Hashino:2016rvx},
discrimination of new physics models
with strongly first-order phase transition
has been discussed at LISA and DECIGO\@.
 The synergy of collider experiments and GW measurement
has been discussed in Refs.~\cite{Hashino:2016rvx, Hashino:2016xoj}.


 There are two $Z_2$-even charged Higgs bosons
$H_1^+$ and $H_2^+$ in our model.
 Since $s_3^+$ does not have the Yukawa interaction,
both of $H_1^+$ and $H_2^+$ decay into fermions
via Yukawa interactions for $\Phi_1$ or $\Phi_2$
through the mixing with $\theta_+$.
 Since we take their Yukawa interactions
to be the same as those in the type-X two Higgs double model,
both of $H_1^+$ and $H_2^+$ decay into $\overline{\tau} \nu$
for the benchmark set in eq.~\eqref{eq:benchmark}.
 See e.g., Refs.~\cite{Kanemura:2011kx, Kanemura:2014dea, Kanemura:2014bqa}
for the prospect of searches for $H^+$
in the the type-X two Higgs double model.
 If such two particles are discovered,
our model might be preferable to the AKS model,
which does not involve the second $Z_2$-even charged Higgs boson.



\section{Conclusions}
\label{sec:concl}

 In this letter,
we have proposed a model
for generating tiny masses of Dirac neutrinos,
where three right-handed neutrinos $\nu_{iR}^{}$
are introduced to the SM
with the lepton number conservation.
 Although Dirac neutrino masses at the tree level
are forbidden by a softly broken $Z_2$ symmetry
in order to avoid unnaturally small coupling constants,
they are generated at the two-loop level,
where neutrino masses are naturally suppressed.
 We have found a benchmark set of parameter values,
which is consistent with neutrino oscillation data,
constraints on charged lepton flavor violations%
~($\ell \to \ellp\gamma$ and $\mu \to \overline{e}ee$)
and the violation of the lepton universality%
~($\ell \to \ellp\nu\overline{\nu}$).
 Masses of new particles can be less than the TeV-scale
which can be probed at the current and future collider experiments,
and unnaturally small coupling constants are not required.
 For the benchmark set,
the dark matter candidate
is the complex scalar ${\mathcal H}^0$
stabilized by the unbroken $Z_2$ symmetry,
which is not imposed by hand
but arises due to assignments of
conserving lepton numbers to new fields.
 We have shown that
the abundance and
the spin-independent scattering cross section
of the dark matter for the benchmark set
are consistent with experimental constraints.
 In addition,
the second $\SU(2)_L$-doublet scalar field
can provide the source of the CP violation in the scalar potential
in principle.
 It has been shown that
the strong first-order phase transition
for the electroweak symmetry breaking can be realized in our model
as required for the electroweak baryogenesis scenario.
 Then,
the deviation of
the coupling constant for the $hhh$ interaction
from the value in the SM
is more than about $20\,\%$,
which can be tested at the future linear collider.
 Notice that
$Z_2$-odd fields~(including the dark matter candidate)
and the second $\SU(2)_L$-doublet scalar field%
~(preferred for the electroweak baryogenesis)
in our model
are essential to generate neutrino masses.
 If one of them is removed from our model,
neutrino masses are not generated.
 Therefore,
our model can be a simultaneous
solution for three problems,
namely non-zero neutrino masses,
the dark matter candidate,
and the baryon asymmetry of the Universe.


\begin{acknowledgments}
This work was supported, in part,
by Grant-in-Aid for Scientific Research on Innovative Areas,
the Ministry of Education, Culture, Sports, Science and Technology,
No.~16H06492, and 
Grant H2020-MSCA-RISE-2014 no.~645722 (Non Minimal Higgs).
\end{acknowledgments}

\appendix
\section{Loop Functions}
\label{sec:app-A}

 The two-loop function $I_{\ell a}$ in eq.~\eqref{eq:mD}
is defined as
\begin{eqnarray}
I_{\ell a}
&\equiv&
 \int\frac{ d^4p }{ (2\pi)^4 }
 \int\frac{ d^4q }{ (2\pi)^4 }
 \frac{1}{ p^2 - m_{H_1^+}^2 }
 \frac{1}{ p^2 - m_{H_2^+}^2 }
 \frac{p^\mu}{ p^2 - m_\ell^2 }
 \frac{1}{ q^2 - m_{{\mathcal H}^0}^2 }
 \frac{q_\mu}{ q^2 - m_{\psi_a}^2 }
 \frac{1}{ (p-q)^2 - m_{{\mathcal H}^+}^2 }
\nonumber\\
&=&
 \frac{1}{
          (4\pi)^4
          \bigl( m_{H_2^+}^2 - m_{H_1^+}^2 \bigr)
          \bigl( m_{\psi_a}^2 - m_{{\mathcal H}_0}^2 \bigr)
         }
\nonumber\\*
&&\hspace*{0mm}
 \int_0^1\!\! dx\,
 x
 \Biggl[
  \frac{1}{ m_\ell^2 - m_{H_2^+}^2 }
  \Bigl\{
   f( m_\ell^2 )
   -
   f( m_{H_2^+}^2 )
  \Bigr\}
  -
  \frac{1}{ m_\ell^2 - m_{H_1^+}^2 }
  \Bigl\{
   f( m_\ell^2 )
   -
   f( m_{H_1^+}^2 )
  \Bigr\}
 \Biggr] ,
\end{eqnarray}
where we used
\begin{eqnarray}
f( m^2 )
&\equiv&
 m^4
 \Bigl\{
  \text{Li}_2 \bigl( z_1(m^2) \bigr)
  - \text{Li}_2 \bigl( z_2(m^2) \bigr)
 \Bigr\} ,
\\
%
%
z_1 (m^2)
&\equiv&
 1
 - \frac{1}{ m^2 x(1-x) }
   \Bigl\{
    m_{{\mathcal H}^0}^2
    + x ( m_{{\mathcal H}^+}^2 - m_{{\mathcal H}^0}^2 )
   \Bigr\} ,
\\
%
%
z_2 (m^2)
&\equiv&
 1
 - \frac{1}{ m^2 x(1-x) }
   \Bigl\{
    m_{\psi_a}^2
    + x ( m_{{\mathcal H}^+}^2 - m_{\psi_a}^2 )
   \Bigr\} ,
\\
%
%
\text{Li}_2(z)
&=&
  - \int_0^z dt
    \frac{1}{\,t\,}
    \ln( 1 - t ) .
\end{eqnarray}
 Ignoring $m_\ell^{}$,
the function can be simplified as
\begin{eqnarray}
I_{\ell a}
=
 I_a
&\equiv&
 \frac{1}{
          (4\pi)^4
          \bigl( m_{H_2^+}^2 - m_{H_1^+}^2 \bigr)
          \bigl( m_{\psi_a}^2 - m_{{\mathcal H}_0}^2 \bigr)
         }
 \int_0^1\!\! dx\,
 x
 \Biggl[
  \frac{f( m_{H_2^+}^2 )}{ m_{H_2^+}^2 }
  -
  \frac{f( m_{H_1^+}^2 )}{ m_{H_1^+}^2 }
 \Biggr] .
\end{eqnarray}

 Loop functions for box diagrams for
$\ell_4(p_4) \to \overline{\ell_3}(-p_3)\, \ell_2(-p_2)\, \ell_1(-p_1)$
are defined as
\begin{eqnarray}
(I^{\text{box}}_1)_{ba}
&\equiv&
 \frac{i^4}{\,4\,}
 \int \frac{d^4 k}{(2\pi)^4}
 \frac{ k^\mu }{ k^2 - m_{\psi_b}^2 }
 \frac{1}{ (k+p_1^{})^2 - m_{{\mathcal H}^+}^2 }
\nonumber\\*
&&\hspace*{25mm}{}
 \times
 \frac{ ( k + p_1^{} + p_3^{} )_\mu }{ (k+p_1^{}+p_3^{})^2 - m_{\psi_a}^2 }
 \frac{1}{ (k+p_1^{}+p_2^{}+p_3^{})^2 - m_{{\mathcal H}^+}^2 }
\nonumber\\
&\simeq&
 -
 \frac{ i }{ 4 (4\pi)^2 }
 \Biggl[
  \frac{ m_{{\mathcal H}^+}^2 }
       {
        ( m_{\psi_a}^2 - m_{{\mathcal H}^+}^2 )
        ( m_{\psi_b}^2 - m_{{\mathcal H}^+}^2 )
       }
  +
  \frac{ m_{\psi_a}^4 }
       {
        ( m_{\psi_b}^2 - m_{\psi_a}^2 )
        ( m_{{\mathcal H}^+}^2 - m_{\psi_a}^2 )^2
       }
  \ln\frac{ m_{{\mathcal H}^+}^2 }{ m_{\psi_a}^2 }
\nonumber\\*
&&\hspace*{25mm}{}
  +
  \frac{ m_{\psi_b}^4 }
       {
        ( m_{\psi_a}^2 - m_{\psi_b}^2 )
        ( m_{{\mathcal H}^+}^2 - m_{\psi_b}^2 )^2
       }
  \ln\frac{ m_{{\mathcal H}^+}^2 }{ m_{\psi_b}^2 }
 \Biggr] ,
\end{eqnarray}
%
%
\begin{eqnarray}
(I^{\text{box}}_2)_{ba}
&\equiv&
 i^4
 \int \frac{d^4 k}{(2\pi)^4}
 \frac{1}{ k^2 - m_{\psi_b}^2 }
 \frac{1}{ (k+p_3^{})^2 - m_{{\mathcal H}^+}^2 }
\nonumber\\*
&&\hspace*{25mm}{}
 \times
 \frac{1}{ (k+p_1^{}+p_3^{})^2 - m_{\psi_a}^2 }
 \frac{1}{ (k+p_1^{}+p_2^{}+p_3^{})^2 - m_{{\mathcal H}^+}^2 }
\nonumber\\
&\simeq&
 -
 \frac{ i }{ (4\pi)^2 }
 \Biggl[
  \frac{ 1 }
       {
        ( m_{\psi_a}^2 - m_{{\mathcal H}^+}^2 )
        ( m_{\psi_b}^2 - m_{{\mathcal H}^+}^2 )
       }
  +
  \frac{ m_{\psi_a}^2 }
       {
        ( m_{\psi_b}^2 - m_{\psi_a}^2 )
        ( m_{{\mathcal H}^+}^2 - m_{\psi_a}^2 )^2
       }
  \ln\frac{ m_{{\mathcal H}^+}^2 }{ m_{\psi_a}^2 }
\nonumber\\*
&&\hspace*{25mm}{}
  +
  \frac{ m_{\psi_b}^2 }
       {
        ( m_{\psi_a}^2 - m_{\psi_b}^2 )
        ( m_{{\mathcal H}^+}^2 - m_{\psi_b}^2 )^2
       }
  \ln\frac{ m_{{\mathcal H}^+}^2 }{ m_{\psi_b}^2 }
 \Biggr] .
\end{eqnarray}
 For $m_\psi^{} \equiv m_{\psi_a}^{} = m_{\psi_b}^{}$,
these functions are simplified as
\begin{eqnarray}
(I^{\text{box}}_1)_{ba}
&\simeq&
  -
  \frac{ i }{ 4 (4\pi)^2 }
  \Biggl\{
   \frac{ m_\psi^2 + m_{{\mathcal H}^+}^2 }
        { ( m_\psi^2 - m_{{\mathcal H}^+}^2 )^2 }
   +
   \frac{ 2 m_\psi^2 m_{{\mathcal H}^+}^2 }
        { ( m_\psi^2 - m_{{\mathcal H}^+}^2 )^3 }
   \ln\frac{ m_{{\mathcal H}^+}^2 }{ m_\psi^2 }
  \Biggr\} ,\\
%
%
(I^{\text{box}}_2)_{ba}
&\equiv&
  -
  \frac{ i }{ (4\pi)^2 }
  \Biggl\{
   \frac{2}{ ( m_\psi^2 - m_{{\mathcal H}^+}^2 )^2 }
   +
   \frac{ m_\psi^2 + m_{{\mathcal H}^+}^2 }
        { ( m_\psi^2 - m_{{\mathcal H}^+}^2 )^3 }
   \ln\frac{ m_{{\mathcal H}^+}^2 }{ m_\psi^2 }
  \Biggr\} .
\end{eqnarray}

\section{Annihilation Cross Section}
\label{sec:app-B}

 The scalar dark matter ${\mathcal H}^0$
can annihilate into $b\overline{b}$
via the tree-level diagram in Fig.~\ref{fig:pair}(a).
 We obtain
\begin{eqnarray}
&&
\sigma({\mathcal H}^0 {\mathcal H}^{0\ast} \to h(H) \to b\overline{b} )\,
v_\rel^{}
\nonumber\\*
&&\hspace*{10mm}
\simeq
 \frac{ 3 m_b^2 }{ 4 \pi v_2^2  }
 \Bigl( 1 - \frac{ m_b^2 }{ m_{{\mathcal H}^0}^2 } \Bigr)^{\frac{3}{2}}
 \Biggl|
  \frac{ \lambda_{h {\mathcal H}^0 {\mathcal H}^0} \cos\alpha}
       { 4 m_{{\mathcal H}^0}^2 - m_h^2 + i m_h^{} \Gamma_h }
  +
  \frac{ \lambda_{H {\mathcal H}^0 {\mathcal H}^0} \sin\alpha }
       { 4 m_{{\mathcal H}^0}^2 - m_H^2 + i m_H^{} \Gamma_H }
 \Biggr|^2 ,
\end{eqnarray}
where
\begin{eqnarray}
 \begin{pmatrix}
  \lambda_{H {\mathcal H}^0 {\mathcal H}^0}\\
  \lambda_{h {\mathcal H}^0 {\mathcal H}^0}
 \end{pmatrix}
 \equiv
  \begin{pmatrix}
   \cos\alpha & \sin\alpha\\
   -\sin\alpha & \cos\alpha
  \end{pmatrix}
  \begin{pmatrix}
   \lambda_{\phi1 s0}\, v_1\\
   \lambda_{\phi2 s0}\, v_2
  \end{pmatrix} .
\end{eqnarray}
 We used $\Gamma_h = 4.07\times 10^{-3}\,\GeV$
for the total width of the discovered Higgs boson~($m_h = 125\,\GeV$),
and the total width of $H$ is calculated as
\begin{eqnarray}
\Gamma_H
=
 \frac{ m_H^{} m_b^2 }{ 8\pi v^2 } \xi_{Hd}^2
 \left( 1 - \frac{ 4 m_b^2 }{ m_H^2 } \right)^{3/2}
 +
 \frac{ m_H^{} m_\tau^2 }{ 8\pi v^2 } \xi_{H\ell}^2
 \left( 1 - \frac{ 4 m_\tau^2 }{ m_H^2 } \right)^{3/2} ,
\end{eqnarray}
where we take
$\xi_{Hd}^{} = \sin\alpha/\sin\beta$ and
$\xi_{H\ell}^{} = \cos\alpha/\cos\beta$.

 There is another tree-level diagram~(Fig.~\ref{fig:pair}(b))
mediated by $\psi_R^0$ with
the Yukawa coupling matrix $Y_\psi^0$.
 The cross section is calculated as
\begin{eqnarray}
&&
\sigma({\mathcal H}^0 {\mathcal H}^{0\ast} \to \nu_R^{}\overline{\nu_R^{}} )\,
v_\rel^{}
\simeq
 \frac{ m_{{\mathcal H}^0}^2 v_\rel^2  }
      { 48 \pi }
  \sum_{a, b}
  \frac{ \left| ( Y_\psi^{0\dagger} Y_\psi^0 )_{ab} \right|^2  }
       {
        \left\{ m_{{\mathcal H}^0}^2 + m_{\psi_a}^2 \right\}
        \left\{ m_{{\mathcal H}^0}^2 + m_{\psi_b}^2 \right\}
       } .
\end{eqnarray}

 The annihilation into a pair of photons
is possible via one-loop diagrams in Figs.~\ref{fig:pair}(c)-(f)
without using Yukawa interactions.
 Contributions of Figs.~\ref{fig:pair}(e) and (f)
are completely cancelled by each other,
and then we have
\begin{eqnarray}
&&
\sigma({\mathcal H}^0 {\mathcal H}^{0\ast} \to \gamma\gamma )\,
v_\rel
\nonumber\\*
&&\hspace*{10mm}
\simeq
 \frac{ 4 \pi \alpha_\EM^2 }
      { (16\pi^2)^2 m_{{\mathcal H}^0}^2 }
 \Biggl[
  \sum_i
    (\mu_{3i}^\prime)^2 
    \Biggl\{
     \frac{ 2 m_{H_i^+}^2 m_{{\mathcal H}^+}^2 }
          { ( m_{H_i^+}^2 - m_{{\mathcal H}^+}^2 )^3 }
     \ln\frac{ m_{H_i^+}^2 }{ m_{{\mathcal H}^+}^2 }
     -
     \frac{ m_{H_i^+}^2 + m_{{\mathcal H}^+}^2 }
          { ( m_{H_i^+}^2 - m_{{\mathcal H}^+}^2 )^2 }
    \Biggr\}
 \Biggr]^2,
\end{eqnarray}
where $\mu_{31}^\prime \equiv \mu_3^\prime \sin\theta_+$
and $\mu_{32}^\prime \equiv \mu_3^\prime \cos\theta_+$.

 The annihilation into a pair of charged leptons
can be caused not only by the tree-level diagram
in Fig.~\ref{fig:pair}(a) with the usual Yukawa interaction
but also by the one-loop diagrams
in Fig.~\ref{fig:pair}(g) and \ref{fig:pair}(h)
with new Yukawa coupling matrix $Y_\psi^+$.
 We obtain
\begin{eqnarray}
\sigma( {\mathcal H}^0 {\mathcal H}^{0\ast} \to \ell \overline{\ellp} )\,
v_\rel
\simeq
 \frac{1}{8\pi}
 \left\{
  ( m_\ell^2 + m_{\ellp}^2 ) |A_{\ell\ellp}|^2
  +
  \frac{2}{\,3\,} m_{{\mathcal H}^0}^2
  |(B_\text{box})_{\ell\ellp}|^2 v_\rel^2
 \right\} ,
\end{eqnarray}
where the following formulae are used:
\begin{eqnarray}
A_{\ell\ellp}
&\equiv&
 (A_\text{tree})_\ell\, \delta_{\ell\ellp}
 + (A_\text{tri})_{\ell\ellp}
 + (A_\text{box})_{\ell\ellp} ,
\\
%
%
(A_\text{tree})_\ell
&\equiv&
 \frac{i}{v_1}
 \Biggl\{
  \frac{ \lambda_{h{\mathcal H}^0{\mathcal H}^0}^{} \sin\alpha }
       { 4 m_{{\mathcal H}^0}^2 - m_h^2 + i m_h \Gamma_h }
  -
  \frac{ \lambda_{H{\mathcal H}^0{\mathcal H}^0}^{} \cos\alpha }
       { 4 m_{{\mathcal H}^0}^2 - m_H^2 + i m_H \Gamma_H }
 \Biggr\} ,
\\
%
%
(A_\text{tri})_{\ell\ellp}
&\equiv&
 - \frac{ i \lambda_{s0s2}^{} }{ 2 (4\pi)^2 }
 \sum_{a}
  (Y_\psi^+)_{\ell a}^\ast (Y_\psi^+)_{a\ellp}^T
 \Biggl\{
  \frac{ m_{\psi_a}^4 }
       { ( m_{{\mathcal H}^+}^2 - m_{\psi_a}^2 )^3 }
  \ln\frac{ m_{{\mathcal H}^+}^2 }{ m_{\psi_a}^2 }
  +
  \frac{ m_{{\mathcal H}^+}^2 - 3 m_{\psi_a}^2 }
       { 2 ( m_{{\mathcal H}^+}^2 - m_{\psi_a}^2 )^2 }
 \Biggr\} ,
\\
%
%
(A_\text{box})_{\ell\ellp}
&\equiv&
 \frac{ i }{ (4\pi)^2 }
 \sum_{a,i}
  (Y_\psi^+)_{\ell a}^\ast (Y_\psi^+)_{a\ellp}^T
  (\mu_{3i}^\prime)^2
\nonumber\\*
&&\hspace*{20mm}{}
 \Biggl\{
  -
  \frac{
        m_{H_i^+}^2 ( m_{H^+}^2 - 2 m_{\psi_a}^2 )
       }
       {
        4
        ( m_{{\mathcal H}^+}^2 - m_{H_i^+}^2 )^2
        ( m_{\psi_a}^2 - m_{H_i^+}^2 )^2
       }
  \ln\frac{ m_{H_i^+}^2 }{ m_{\psi_a}^2 }
\nonumber\\*
&&\hspace*{25mm}{}
  -
  \frac{
        2 m_{\psi_a}^4 m_{H_i^+}^2
        -
        3 m_{\psi_a}^2 m_{{\mathcal H}^+}^4
        +
        m_{{\mathcal H}^+}^6
       }
       {
        4
        ( m_{\psi_a}^2 - m_{{\mathcal H}^+}^2 )^3
        ( m_{H_i^+}^2 - m_{{\mathcal H}^+}^2  )^2
       }
  \ln\frac{ m_{{\mathcal H}^+}^2 }{ m_{\psi_a}^2 }
\nonumber\\*
&&\hspace*{25mm}{}
  +
  \frac{
        2 m_{\psi_a}^4
        -
        3 m_{\psi_a}^2 m_{H_i^+}^2
        +
        m_{{\mathcal H}^+}^2 m_{H_i^+}^2
       }
       {
        4
        ( m_{\psi_a}^2 - m_{{\mathcal H}^+}^2 )^2
        ( m_{\psi_a}^2 - m_{H_i^+}^2 )
        ( m_{{\mathcal H}^+}^2 - m_{H_i^+}^2 )
       }
 \Biggr\} ,
\\
%
%
(B_\text{box})_{\ell\ellp}
&\equiv&
 \frac{ i }{ (4\pi)^2 }
 \sum_{a,i}
  (Y_\psi^+)_{\ell a}^\ast (Y_\psi^+)_{a\ellp}^T
  (\mu_{3i}^\prime)^2
\nonumber\\*
&&\hspace*{20mm}{}
 \Biggl\{
  - \frac{
          m_{H_i^+}^2
          (
           - 2 m_{\psi_a}^2 m_{{\mathcal H}^+}^2
           + m_{{\mathcal H}^+}^2 m_{H_i^+}^2
           + m_{H_i^+}^4
          )
         }
         {
          4
          ( m_{\psi_a}^2 - m_{H_i^+}^2 )^2
          ( m_{{\mathcal H}^+}^2 - m_{H_i^+}^2 )^3
         }
   \ln\frac{ m_{H_i^+}^2 }{ m_{\psi_a}^2 }
\nonumber\\*
&&\hspace*{25mm}{}
  -
  \frac{
        m_{{\mathcal H}^+}^2
        (
         - 2 m_{\psi_a}^2 m_{H_i^+}^2
         + m_{{\mathcal H}^+}^2 m_{H_i^+}^2
         + m_{{\mathcal H}^+}^4
        )
       }
       {
        4
        ( m_{\psi_a}^2 - m_{{\mathcal H}^+}^2 )^2
        ( m_{H_i^+}^2 - m_{{\mathcal H}^+}^2 )^3
       }
  \ln\frac{ m_{{\mathcal H}^+}^2 }{ m_{\psi_a}^2 }
\nonumber\\*
&&\hspace*{25mm}{}
  +
  \frac{
        m_{\psi_a}^2
        ( m_{{\mathcal H}^+}^2 + m_{H_i^+}^2 )
        -
        2 m_{{\mathcal H}^+}^2 m_{H_i^+}^2
       }
       {
        4
        ( m_{\psi_a}^2 - m_{{\mathcal H}^+}^2 )
        ( m_{\psi_a}^2 - m_{H_i^+}^2 )
        ( m_{{\mathcal H}^+}^2 - m_{H_i^+}^2 )^2
       }
 \Biggr\}
\end{eqnarray}

 The self-annihilation of ${\mathcal H}^0$
into two $\overline{\nu_R^{}}$
is also possible via the diagram in Fig.~\ref{fig:self},
which results in
\begin{eqnarray}
\sigma({\mathcal H}^0 {\mathcal H}^0
\to \overline{\nu_R^{}}\, \overline{\nu_R^{}} )\, v_\rel^{}
\simeq
 \sum_{a, b}
 \frac{
       \left\{ ( Y_\psi^{0\dagger} Y_\psi^0 )_{ab} \right\}^2
       m_{\psi_a}^{} m_{\psi_b}^{}
      }
      {
       4 \pi
       ( m_{{\mathcal H}^0}^2 + m_{\psi_a}^2 )
       ( m_{{\mathcal H}^0}^2 + m_{\psi_b}^2 )
      } .
\end{eqnarray}

\section{Field-Dependent Masses}
\label{sec:app-C}

Field-dependent masses $\wt{m}_i^{}(\varphi)$ of particles $i$
are given by
\begin{eqnarray}
\wt{m}_h^2(\varphi)
&=&
 \frac{3 m_h^2}{2}
 \left( \frac{\varphi^2}{v^2} - \frac{1}{\,3\,} \right) ,
\\
%
%
\wt{m}_{G^0}^2(\varphi)
&=&
 \wt{m}_{G^\pm}^2(\varphi)
=
 \frac{m_h^2}{2} \left( \frac{\varphi^2}{v^2} - 1 \right) ,
\\
%
%
\wt{m}_{H}^2(\varphi)
&=&
 \left( m_H^2 - \frac{2 m_{12}^2}{\sin(2\beta)} + \frac{m_h^2}{2} \right)
 \frac{\varphi^2}{v^2}
 +
 \left( \frac{2 m_{12}^2}{\sin(2\beta)} - \frac{m_h^2}{2} \right) ,
\\
%
%
\wt{m}_{A}^2(\varphi)
&=&
 \left( m_A^2 - \frac{2 m_{12}^2}{\sin(2\beta)} + \frac{m_h^2}{2} \right)
 \frac{\varphi^2}{v^2}
 +
 \left( \frac{2 m_{12}^2}{\sin(2\beta)} - \frac{m_h^2}{2} \right) ,
\\
%
%
\wt{m}_{\mathcal{H}^0}^2(\varphi)
&=&
 \left( m_{\mathcal{H}^0}^2 - m_{s0}^2 \right)
 \frac{\varphi^2}{v^2}
 +
 m_{s0}^2 ,
\\
%
%
\wt{m}_{\mathcal{H}^\pm}^2(\varphi)
&=&
 \left( m_{\mathcal{H}^\pm}^2 - m_{s2}^2 \right)
 \frac{\varphi^2}{v^2}
 +
 m_{s2}^2 ,
\\
%
%
\wt{m}_{H^\pm_1}^2(\varphi)
&=&
 \frac{1}{\,2\,}
 \left\{
  (\wt{M}_{H^+}^{\prime 2})_{11} + (\wt{M}_{H^+}^{\prime 2})_{22}
  - \sqrt{
          \bigl(
           (\wt{M}_{H^+}^{\prime 2})_{22}
           - (\wt{M}_{H^+}^{\prime 2})_{11}
          \bigr)^2
           - 4 (\wt{M}_{H^+}^{\prime 2})_{12}^2
         }\,
 \right\} ,
\\
%
%
\wt{m}_{H^\pm_2}^2(\varphi)
&=&
 \frac{1}{\,2\,}
 \left\{
  (\wt{M}_{H^+}^{\prime 2})_{11} + (\wt{M}_{H^+}^{\prime 2})_{22}
  + \sqrt{
          \bigl(
           (\wt{M}_{H^+}^{\prime 2})_{22}
           - (\wt{M}_{H^+}^{\prime 2})_{11}
          \bigr)^2
           - 4 (\wt{M}_{H^+}^{\prime 2})_{12}^2
         }\,
 \right\} ,
\\
%
%
(\wt{M}_{H^+}^{\prime 2})_{11}
&=&
 \left(
  m_{H^\pm_1}^2 \cos^2\theta_+
  + m_{H^\pm_2}^2 \sin^2\theta_+
  - \frac{2 m_{12}^2}{\sin(2\beta)}
  + \frac{m_h^2}{2}
 \right) \frac{\varphi^2}{v^2}
 +
 \frac{2 m_{12}^2}{\sin(2\beta)}
 -
 \frac{m_h^2}{2} ,
\\
%
%
(\wt{M}_{H^+}^{\prime 2})_{22}
&=&
 \left(
  m_{H^\pm_1}^2 \sin^2\theta_+
  + m_{H^\pm_2}^2 \cos^2\theta_+
  - m_{s3}^2
 \right) \frac{\varphi^2}{v^2}
 +
 m_{s3}^2 ,
\\
%
%
(\wt{M}_{H^+}^{\prime 2})_{12}
&=&
 \frac{\varphi}{2v}
 (m_{H^\pm_1}^2 - m_{H^\pm_2}^2 ) \sin(2\theta_+),
\\
%
%
\wt{m}_t^2(\varphi)
&=&
 m_t^2 \frac{\varphi^2}{v^2} ,
\\
%
%
\wt{m}_W^2(\varphi)
&=&
 m_W^2 \frac{\varphi^2}{v^2} ,
\\
%
%
\wt{m}_Z^2(\varphi)
&=&
 m_Z^2 \frac{\varphi^2}{v^2} ,
\end{eqnarray}
where we take
$2 m_{12}^2/\sin(2\beta) = (200\,\GeV)^2$,
$m_{s2}^2 = 0$,
and $m_{s3}^2 = (300\,\GeV)^2$.
 Notice that
$m_{s0}^2 \simeq (61\,\GeV)^2$ is obtained with
eqs.~\eqref{eq:mclH}, \eqref{eq:benchmark}, and \eqref{eq:benchmark-2}.


\section{Thermal Masses}
\label{sec:app-D}

 Field-dependent masses are
thermally corrected at a finite temperature
by contributions of ring diagrams.
 We only focus on the leading terms of $\mathcal{O}(T^2)$%
~\cite{Carrington:1991hz, Sagunski:2016dmz}.
 At a finite temperature $T$,
field-dependent masses $\wt{m}_i^{}(\varphi, T)$
of $Z_2$-odd scalar particles%
~($i = {\mathcal H}^0$, ${\mathcal H}^\pm$)
are given by
\begin{eqnarray}
\wt{m}_{{\mathcal H}^0}^2(\varphi, T)
&=&
 \wt{m}_{{\mathcal H}^0}^2(\varphi)
 +
 \frac{T^2}{12}
 \Bigl\{
  2( \lambda_{\phi1 s0} + \lambda_{\phi2 s0})
  + 4\lambda_{s0}
  + \lambda_{s0s2} + \lambda_{s0s3}
 \Bigr\} ,
\\
%
%
\wt{m}_{{\mathcal H}^\pm}^2(\varphi, T)
&=&
 \wt{m}_{{\mathcal H}^\pm}^2(\varphi)
 +
 \frac{T^2}{12}
 \Bigl\{
  2( \lambda_{\phi1 s2} + \lambda_{\phi2 s2})
  + 4\lambda_{s2}
  + \lambda_{s0s2} + \lambda_{s2s3}
 \Bigr\} .
\end{eqnarray}

 Field-dependent squared masses $\wt{m}_i^2(\varphi, T)$
of $Z_2$-even scalar particles%
~($i = h, H, G^0, A^0, G^+, H_1^+$, and $H_2^+$)
at a finite temperature
are given as eigenvalues of the following matrices
($\wt{M}_H^2(\varphi, T)$ for CP-even ones,
$\wt{M}_A^2(\varphi, T)$ for CP-odd ones,
and $\wt{M}_{H^+}^2(\varphi, T)$ for charged ones)
\begin{eqnarray}
\wt{M}_H^2(\varphi, T)
&=&
 \begin{pmatrix}
  \cos\beta & -\sin\beta\\
  \sin\beta & \cos\beta
 \end{pmatrix}
 \begin{pmatrix}
  \wt{m}_h^2(\varphi) & 0\\
  0 & \wt{m}_H^2(\varphi)
 \end{pmatrix}
 \begin{pmatrix}
  \cos\beta & \sin\beta\\
  -\sin\beta & \cos\beta
 \end{pmatrix}
\nonumber\\*
&&\hspace*{80mm}{}
 +
 \begin{pmatrix}
  \Pi_{\phi_1}(T) & 0\\
  0 & \Pi_{\phi_2}(T)
 \end{pmatrix} ,
\\
%
%
\wt{M}_A^2(\varphi, T)
&=&
 \begin{pmatrix}
  \cos\beta & -\sin\beta\\
  \sin\beta & \cos\beta
 \end{pmatrix}
 \begin{pmatrix}
  \wt{m}_{G^0}^2(\varphi) & 0\\
  0 & \wt{m}_A^2(\varphi)
 \end{pmatrix}
 \begin{pmatrix}
  \cos\beta & \sin\beta\\
  -\sin\beta & \cos\beta
 \end{pmatrix}
\nonumber\\*
&&\hspace*{80mm}{}
 +
 \begin{pmatrix}
  \Pi_{\phi_1}(T) & 0\\
  0 & \Pi_{\phi_2}(T)
 \end{pmatrix} ,
\\
%
%
\wt{M}_{H^+}^2(\varphi, T)
&=&
 \begin{pmatrix}
  \cos\beta & \ - \sin\beta \cos\theta_+ & \sin\beta \sin\theta_+\\
  \sin\beta & \cos\beta \cos\theta_+ & - \cos\beta \sin\theta_+\\
  0 & \sin\theta_+ & \cos\theta_+
 \end{pmatrix}
 \begin{pmatrix}
  \wt{m}_{G^+}^2(\varphi) & 0 & 0\\
  0 & \wt{m}_{H_1^+}^2(\varphi) & 0\\
  0 & 0 & \wt{m}_{H_2^+}^2(\varphi)
 \end{pmatrix}
\nonumber\\*
&&\hspace*{8mm}{}
 \begin{pmatrix}
  \cos\beta & \sin\beta & 0\\
  - \sin\beta \cos\theta_+ & \cos\beta \cos\theta_+ & \sin\theta_+\\
  \sin\beta \sin\theta_+ & - \cos\beta \sin\theta_+ & \cos\theta_+
 \end{pmatrix}
 +
 \begin{pmatrix}
  \Pi_{\phi_1}(T) & 0 & 0\\
  0 & \Pi_{\phi_2}(T) & 0\\
  0 & 0 & \Pi_{s_3^+}(T)
 \end{pmatrix} ,
\nonumber\\
\end{eqnarray}
where thermal masses are given by
\begin{eqnarray}
\Pi_{\phi_1}(T)
&=&
 T^2
 \left\{
  \frac{3 g^2 + g^{\prime 2}}{16}
  + \frac{y_\tau^{}}{12}
  + \frac{1}{12}
    \Bigl(
     3\lambda_1 + 2\lambda_3 + \lambda_4
     + \lambda_{\phi1 s0} + \lambda_{\phi1 s2} +\lambda_{\phi1 s3}
    \Bigr)
 \right\} ,
\\
%
%
\Pi_{\phi_2}(T)
&=&
 T^2
 \left\{
  \frac{3 g^2 + g^{\prime 2}}{16}
  + \frac{y_t^{} + y_b^{}}{12}
  + \frac{1}{12}
    \Bigl(
     3\lambda_2 + 2\lambda_3 + \lambda_4
     + \lambda_{\phi2 s0} + \lambda_{\phi2 s2} +\lambda_{\phi2 s3}
    \Bigr)
 \right\} ,
\\
%
%
\Pi_{s_3^+}(T)
&=&
 \frac{T^2}{12}
 \Bigl\{
  2( \lambda_{\phi1 s3} + \lambda_{\phi2 s3})
  + 4\lambda_{s3}
  + \lambda_{s0s3} + \lambda_{s2s3}
 \Bigr\} .
\end{eqnarray}

 The field-dependent mass $\wt{m}_W^2(\varphi, T)$
of the $W$ boson is given by
\begin{eqnarray}
\wt{m}_W^2(\varphi, T)
&=&
 \wt{m}_W^2(\varphi) + 2 g^2 T^2 .
\end{eqnarray}
 On the other hand,
$\wt{m}_i^2(\varphi, T)$ for $i = Z$ and $\gamma$ are
obtained as eigenvalues of the following matrix:
\begin{eqnarray}
\wt{M}_Z^2(\varphi, T)
&=&
 \frac{\varphi^2}{4}
 \begin{pmatrix}
  g^2 & g g^\prime\\
  g g^\prime & g^{\prime2}
 \end{pmatrix}
 +
 2 T^2
 \begin{pmatrix}
  g^2 & 0\\
  0 & g^{\prime 2}
 \end{pmatrix} .
\end{eqnarray}

 In the calculation for $\varphi_c/T_c$,
we take
$\lambda_{\phi1 s0}^{} = 0.02$
and $\lambda_{\phi2 s0}^{} = 0.005$
in eq.~\eqref{eq:benchmark-2}
as well as
$\lambda_{\phi1 s2} = \lambda_{\phi1 s3} = 0.1$,
$\lambda_{s0s2} = \lambda_{s0s3} = \lambda_{s2s3} = 0$,
and $\lambda_{s0} = \lambda_{s2} = \lambda_{s3} = 0$.
 The other coupling constants in the scalar potential $V$
are determined by
\begin{eqnarray}
\lambda_1
&=&
 \frac{1}{v^2 \cos^2\beta}
 \left(
  - m_{12}^2 \tan\beta
  + m_h^2 \sin^2\alpha
  + m_H^2 \cos^2\alpha
 \right) ,
\\
%
%
\lambda_2
&=&
 \frac{1}{v^2 \sin^2\beta}
 \left(
  - m_{12}^2 \cot\beta
  + m_h^2 \cos^2\alpha
  + m_H^2 \sin^2\alpha
 \right) ,
\\
%
%
\lambda_3
&=&
 \frac{1}{v^2}
 \left\{
  - \frac{2 m_{12}^2}{\sin(2\beta)}
  + \frac{ \sin(2\alpha) }{ \sin(2\beta) }
    ( m_H^2 - m_h^2 )
  + 2 m_{H_1^+}^2 \cos^2\theta_+
  + 2 m_{H_2^+}^2 \sin^2\theta_+
 \right\} ,
\\
%
%
\lambda_4
&=&
 \frac{1}{v^2}
 \left(
  \frac{2 m_{12}^2}{\sin(2\beta)}
  + m_A^2
  - 2 m_{H_1^+}^2 \cos^2\theta_+
  - 2 m_{H_2^+}^2 \sin^2\theta_+
 \right) ,
\\
%
%
\lambda_5
&=&
 \frac{1}{v^2}
 \left(
  \frac{2 m_{12}^2}{\sin(2\beta)}
  - m_A^2
 \right) ,
\\
%
%
%
%
\lambda_{\phi2 s2}
&=&
 \frac{1}{v^2 \sin^2\beta}
 \left(
  2 m_{{\mathcal H}^\pm}^2
  - 2 m_{s2}^2
  - \lambda_{\phi1 s2}\, v^2 \cos^2\beta
 \right) ,
\\
%
%
\lambda_{\phi2 s3}
&=&
 \frac{1}{v^2 \sin^2\beta}
 \left(
  2 m_{H_1^\pm}^2 \sin^2\theta_+
  + 2 m_{H_2^\pm}^2 \cos^2\theta_+
  - 2 m_{s3}^2
  - \lambda_{\phi1 s3}\, v^2 \cos^2\beta
 \right) .
\end{eqnarray}
 These values at the benchmark point are
$\lambda_1 \simeq 0.26$,
$\lambda_2 \simeq 0.26$,
$\lambda_3 \simeq 0.27$,
$\lambda_4 \simeq -0.017$,
$\lambda_5 = 0$,
$\lambda_{\phi2 s2} \simeq 4.5$,
and $\lambda_{\phi2 s3} \simeq -0.030$.



\begin{thebibliography}{99}


\bibitem{ref:2012Jul}
%
  G.~Aad {\it et al.} [ATLAS Collaboration],
  Phys.\ Lett.\ B {\bf 716}, 1 (2013);
%
%
  S.~Chatrchyan {\it et al.} [CMS Collaboration],
  Phys.\ Lett.\ B {\bf 716}, 30 (2012).


\bibitem{ref:numass}
  Y.~Fukuda {\it et al.} [Super-Kamiokande Collaboration],
  Phys.\ Rev.\ Lett.\  {\bf 81}, 1562 (1998);
%
%
  Q.~R.~Ahmad {\it et al.} [SNO Collaboration],
  Phys.\ Rev.\ Lett.\  {\bf 89}, 011301 (2002).




\bibitem{Ade:2015xua} 
  P.~A.~R.~Ade {\it et al.} [Planck Collaboration],
  Astron.\ Astrophys.\  {\bf 594}, A13 (2016).



%
\bibitem{ref:seesaw}
  P.~Minkowski,
  Phys.\ Lett.\  B {\bf 67}, 421 (1977);
%
%
  T.~Yanagida,
  Conf.\ Proc.\ C {\bf 7902131}, 95 (1979);
%
%
  Prog.\ Theor.\ Phys.\  {\bf 64}, 1103 (1980);
%
%
  M.~Gell-Mann, P.~Ramond and R.~Slansky,
  Conf.\ Proc.\ C {\bf 790927}, 315 (1979);
%
%
  R.~N.~Mohapatra and G.~Senjanovic,
  Phys.\ Rev.\ Lett.\  {\bf 44}, 912 (1980);
%
%
J.~Schechter and J.~W.~F.~Valle,
Phys.\ Rev.\ D {\bf 22}, 2227 (1980).





\bibitem{Fukugita:1986hr} 
  M.~Fukugita and T.~Yanagida,
  Phys.\ Lett.\ B {\bf 174}, 45 (1986).



\bibitem{ref:Ma}
%
  E.~Ma,
  Phys.\ Rev.\ D {\bf 73}, 077301 (2006);
%
  J.~Kubo, E.~Ma and D.~Suematsu,
  Phys.\ Lett.\ B {\bf 642}, 18 (2006).



\bibitem{ref:AKS}
%
  M.~Aoki, S.~Kanemura and O.~Seto,
  Phys.\ Rev.\ Lett.\  {\bf 102}, 051805 (2009);
%
  Phys.\ Rev.\ D {\bf 80}, 033007 (2009);
%
  M.~Aoki, S.~Kanemura and K.~Yagyu,
  Phys.\ Rev.\ D {\bf 83}, 075016 (2011).



\bibitem{Kuzmin:1985mm} 
  V.~A.~Kuzmin, V.~A.~Rubakov and M.~E.~Shaposhnikov,
  Phys.\ Lett.\  {\bf 155B}, 36 (1985).




\bibitem{Kanemura:2015cca} 
  S.~Kanemura and H.~Sugiyama,
  Phys.\ Lett.\ B {\bf 753}, 161 (2016).



\bibitem{Kanemura:2016ixx} 
  S.~Kanemura, K.~Sakurai and H.~Sugiyama,
  Phys.\ Lett.\ B {\bf 758}, 465 (2016).





\bibitem{Aoki:2009ha} 
  M.~Aoki, S.~Kanemura, K.~Tsumura and K.~Yagyu,
  Phys.\ Rev.\ D {\bf 80}, 015017 (2009).


\bibitem{Barger:1989fj} 
  V.~D.~Barger, J.~L.~Hewett and R.~J.~N.~Phillips,
  Phys.\ Rev.\ D {\bf 41}, 3421 (1990).


\bibitem{Grossman:1994jb} 
  Y.~Grossman,
  Nucl.\ Phys.\ B {\bf 426}, 355 (1994).


\bibitem{ref:THDM-Akeroyd}
  A.~G.~Akeroyd and W.~J.~Stirling,
  Nucl.\ Phys.\ B {\bf 447}, 3 (1995);
%
%
  A.~G.~Akeroyd,
  Phys.\ Lett.\ B {\bf 377}, 95 (1996);
%
%
  J.\ Phys.\ G {\bf 24}, 1983 (1998).





\bibitem{Maki:1962mu} 
  Z.~Maki, M.~Nakagawa and S.~Sakata,
  Prog.\ Theor.\ Phys.\  {\bf 28}, 870 (1962).



\bibitem{Abe:2015awa} 
  K.~Abe {\it et al.} [T2K Collaboration],
  Phys.\ Rev.\ D {\bf 91}, no. 7, 072010 (2015).


\bibitem{An:2016ses} 
  F.~P.~An {\it et al.} [Daya Bay Collaboration],
  arXiv:1610.04802 [hep-ex].


\bibitem{Aharmim:2011vm} 
  B.~Aharmim {\it et al.}  [SNO Collaboration],
  Phys.\ Rev.\ C {\bf 88}, no. 2, 025501 (2013).



\bibitem{Kolb:1990vq} 
  E.~W.~Kolb and M.~S.~Turner,
  ``The Early Universe,''
  Front.\ Phys.\  {\bf 69}, 1 (1990).




\bibitem{DiLuzio:2014bua} 
  L.~Di Luzio and L.~Mihaila,
  JHEP {\bf 1406}, 079 (2014).


\bibitem{Dolan:1973qd} 
  L.~Dolan and R.~Jackiw,
  Phys.\ Rev.\ D {\bf 9}, 3320 (1974).


\bibitem{Carrington:1991hz} 
  M.~E.~Carrington,
  Phys.\ Rev.\ D {\bf 45}, 2933 (1992).


\bibitem{Cline:1996mga} 
  J.~M.~Cline and P.~A.~Lemieux,
  Phys.\ Rev.\ D {\bf 55}, 3873 (1997).






\bibitem{TheMEG:2016wtm} 
  A.~M.~Baldini {\it et al.} [MEG Collaboration],
  Eur.\ Phys.\ J.\ C {\bf 76}, no. 8, 434 (2016).


\bibitem{Aubert:2009ag} 
  B.~Aubert {\it et al.} [BaBar Collaboration],
  Phys.\ Rev.\ Lett.\  {\bf 104}, 021802 (2010).



\bibitem{Baldini:2013ke} 
  A.~M.~Baldini {\it et al.},
  arXiv:1301.7225 [physics.ins-det].



\bibitem{Bellgardt:1987du} 
  U.~Bellgardt {\it et al.} [SINDRUM Collaboration],
  Nucl.\ Phys.\ B {\bf 299}, 1 (1988).


\bibitem{Blondel:2013ia} 
  A.~Blondel {\it et al.},
  arXiv:1301.6113 [physics.ins-det].






\bibitem{Olive:2016xmw} 
  C.~Patrignani {\it et al.} [Particle Data Group],
  Chin.\ Phys.\ C {\bf 40}, no. 10, 100001 (2016).


\bibitem{Aubert:2009qj} 
  B.~Aubert {\it et al.} [BaBar Collaboration],
  Phys.\ Rev.\ Lett.\  {\bf 105}, 051602 (2010).





\bibitem{Ellis:2008hf} 
  J.~R.~Ellis, K.~A.~Olive and C.~Savage,
  Phys.\ Rev.\ D {\bf 77}, 065026 (2008).


\bibitem{ref:y=0}
  H.~Ohki {\it et al.},
  Phys.\ Rev.\ D {\bf 78}, 054502 (2008);
%
  PoS LAT {\bf 2009}, 124 (2009).



\bibitem{Akerib:2016vxi} 
  D.~S.~Akerib {\it et al.} [LUX Collaboration],
  Phys.\ Rev.\ Lett.\  {\bf 118}, no. 2, 021303 (2017).



\bibitem{Akerib:2015cja} 
  D.~S.~Akerib {\it et al.} [LZ Collaboration],
  arXiv:1509.02910 [physics.ins-det].


\bibitem{Aprile:2015uzo} 
  E.~Aprile {\it et al.} [XENON Collaboration],
  JCAP {\bf 1604}, no. 04, 027 (2016).









\bibitem{Khachatryan:2016vau} 
  G.~Aad {\it et al.} [ATLAS and CMS Collaborations],
  JHEP {\bf 1608}, 045 (2016).




\bibitem{Dawson:2013bba} 
  S.~Dawson {\it et al.},
  arXiv:1310.8361 [hep-ex].


\bibitem{Fujii:2015jha} 
  K.~Fujii {\it et al.},
  arXiv:1506.05992 [hep-ex].






\bibitem{Barklow:2015tja} 
  T.~Barklow, J.~Brau, K.~Fujii, J.~Gao, J.~List, N.~Walker and K.~Yokoya,
  arXiv:1506.07830 [hep-ex].




\bibitem{CMS:2015nat} 
  CMS Collaboration [CMS Collaboration],
  CMS-PAS-FTR-15-002.


\bibitem{Fujii:2017ekh} 
  K.~Fujii {\it et al.},
  arXiv:1702.05333 [hep-ph].






\bibitem{Kanemura:2004ch} 
  S.~Kanemura, Y.~Okada and E.~Senaha,
  Phys.\ Lett.\ B {\bf 606}, 361 (2005)
  doi:10.1016/j.physletb.2004.12.004
  [hep-ph/0411354].




\bibitem{Apreda:2001us} 
  R.~Apreda, M.~Maggiore, A.~Nicolis and A.~Riotto,
  Nucl.\ Phys.\ B {\bf 631}, 342 (2002).

\bibitem{Grojean:2006bp} 
  C.~Grojean and G.~Servant,
  Phys.\ Rev.\ D {\bf 75}, 043507 (2007).

\bibitem{Espinosa:2008kw} 
  J.~R.~Espinosa, T.~Konstandin, J.~M.~No and M.~Quiros,
  Phys.\ Rev.\ D {\bf 78}, 123528 (2008).

\bibitem{Caprini:2015zlo} 
  C.~Caprini {\it et al.},
  JCAP {\bf 1604}, no. 04, 001 (2016).

\bibitem{Huber:2015znp} 
  S.~J.~Huber, T.~Konstandin, G.~Nardini and I.~Rues,
  JCAP {\bf 1603}, no. 03, 036 (2016).

\bibitem{Chala:2016ykx} 
  M.~Chala, G.~Nardini and I.~Sobolev,
  Phys.\ Rev.\ D {\bf 94}, no. 5, 055006 (2016).

\bibitem{Kobakhidze:2016mch} 
  A.~Kobakhidze, A.~Manning and J.~Yue,
  arXiv:1607.00883 [hep-ph].








\bibitem{Audley:2017drz} 
  H.~Audley {\it et al.},
  arXiv:1702.00786 [astro-ph.IM].
See also
\verb+https://lisa.nasa.gov/+



\bibitem{Kawamura:2011zz} 
  S.~Kawamura {\it et al.},
  Class.\ Quant.\ Grav.\  {\bf 28}, 094011 (2011).


\bibitem{Corbin:2005ny} 
  V.~Corbin and N.~J.~Cornish,
  Class.\ Quant.\ Grav.\  {\bf 23}, 2435 (2006).




\bibitem{Kakizaki:2015wua} 
  M.~Kakizaki, S.~Kanemura and T.~Matsui,
  Phys.\ Rev.\ D {\bf 92}, no. 11, 115007 (2015).


\bibitem{Hashino:2016rvx} 
  K.~Hashino, M.~Kakizaki, S.~Kanemura and T.~Matsui,
  Phys.\ Rev.\ D {\bf 94}, no. 1, 015005 (2016).


\bibitem{Hashino:2016xoj} 
  K.~Hashino, M.~Kakizaki, S.~Kanemura, P.~Ko and T.~Matsui,
  Phys.\ Lett.\ B {\bf 766}, 49 (2017).






\bibitem{Kanemura:2011kx} 
  S.~Kanemura, K.~Tsumura and H.~Yokoya,
  Phys.\ Rev.\ D {\bf 85}, 095001 (2012).


\bibitem{Kanemura:2014dea} 
  S.~Kanemura, H.~Yokoya and Y.~J.~Zheng,
  Nucl.\ Phys.\ B {\bf 886}, 524 (2014).


\bibitem{Kanemura:2014bqa} 
  S.~Kanemura, K.~Tsumura, K.~Yagyu and H.~Yokoya,
  Phys.\ Rev.\ D {\bf 90}, 075001 (2014).




\bibitem{Sagunski:2016dmz} 
  L.~Sagunski,
  DESY-THESIS-2016-023.
\end{thebibliography}
\end{document}